\documentclass[12pt]{article}
\pdfoutput=1  
 
\usepackage{hyperref}    
\usepackage{amsmath}      
\usepackage{amssymb} 
\usepackage{graphicx}
\usepackage{url}
\usepackage{enumerate}
 
\usepackage{float,wrapfig,color} 

\allowdisplaybreaks
\def\eq#1{(\ref{#1})}
\def\s[#1\s]{\begin{align}\begin{split}#1\end{split}\end{align}}
\def\[#1\]{\begin{align}#1\end{align}}

\newcommand{\mathd}{\mathrm{d}}
\newcommand{\mathe}{\mathrm{e}}

\begin{document}

\begin{titlepage} 

\title{
\hfill\parbox{4cm}{ \normalsize YITP-21-99}\\  
\vspace{1cm} 
Emergence of Lie group symmetric classical spacetimes\\ in canonical tensor model
}

\author{
Taigen Kawano\footnote{taigen.kawano@yukawa.kyoto-u.ac.jp}
and 
Naoki Sasakura\footnote{sasakura@yukawa.kyoto-u.ac.jp}
\\
{\small{\it Yukawa Institute for Theoretical Physics, Kyoto University,}}
\\ {\small{\it  Kitashirakawa, Sakyo-ku, Kyoto 606-8502, Japan}}
}


\maketitle

\begin{abstract} 
We analyze a wave function of a tensor model in the canonical formalism,
when the argument of the wave function takes Lie group invariant or nearby values.
Numerical computations show that there are  two phases, which we call the quantum and the classical phases, respectively. 
In the classical phase, fluctuations are suppressed,  and there emerge configurations
which are discretizations of  the classical geometric spaces invariant under the Lie group symmetries.
This is explicitly demonstrated for the emergence of $S^n\ (n=1,2,3)$ for $SO(n+1)$ symmetries by 
checking the topological and the geometric (Laplacian) properties of the emerging configurations.
The transition between the two phases has the form of splitting/merging of distributions of variables,
resembling a matrix model  counterpart, namely, 
the transition between one-cut and two-cut solutions.
However this resemblance is obscured by a difference of the mechanism of the distribution in our setup from that in the matrix model.
We also discuss this transition as a replica symmetry breaking.
We perform various preliminary studies of the properties of the phases and the transition for such values of the argument.
\end{abstract}
\end{titlepage}

\section{Introduction}
\label{sec:introduction}
Lie groups \cite{Georgi:1999wka}
are ubiquitous in spacetime. Fundamental interactions are disciplined by Lie group gauge symmetries, 
Lorentz symmetry constrains spacetime structure of theories, and de Sitter symmetry is globally realized in the Universe.
Therefore, when we think of emergence of spacetimes in quantum gravity \cite{Reuter:2019byg,Eichhorn:2018yfc,Loll:2019rdj,Rovelli:2014ssa,Surya:2019ndm,Konopka:2006hu,Wolfram:2020jjc,
Trugenberger:2016viw,Akara-pipattana:2021zzy,Ambjorn:1990ge,Sasakura:1990fs,Godfrey:1990dt,Gurau:2009tw,Sasakura:2011sq},  
it would be natural to think of emergence of 
Lie group symmetries at the same time, or even consider them as different aspects of one phenomenon.  

In this paper, we discuss a tensor model in the Hamiltonian formalism from the perspective above.
Tensor models \cite{Ambjorn:1990ge,Sasakura:1990fs,Godfrey:1990dt,Gurau:2009tw}
were originally introduced as a generalization of the matrix model, 
which successfully describes two-dimensional quantum gravity, to higher dimensions.
However, tensor models were shown to suffer from dominance of singular spaces \cite{Bonzom:2011zz,Gurau:2011xp} 
and seem difficult to generate globally extended spaces. 
With a hope to improve the issue, one of the present authors introduced a new type of tensor model with a time
direction \cite{Sasakura:2011sq,Sasakura:2012fb}, which we call canonical tensor model (CTM).
CTM incorporates an analogue of the spacetime diffeomorphism invariance,
which is a fundamental in general relativity, 
by mimicking the structure of the Hamiltonian formalism of general relativity (more precisely, the ADM formalism \cite{Arnowitt:1962hi}).   
The reason why we think introducing time may improve the difficulty comes from 
the success of the causal dynamical triangulation \cite{Ambjorn:2004qm} 
over the the dynamical triangulation in emergence of globally extended spacetimes, 
where the former is a dynamical lattice formulation of quantum gravity with a dime direction, but the latter is one without it.
It would be a highly interesting question whether CTM enjoys a similar success or not.
From the previous studies \cite{Obster:2017dhx,Obster:2017pdq},
 it is known that the wave function $\Psi(P)$ of CTM has peaks at Lie group invariant $P$'s,
where $P$ denotes one of the canonically conjugate pair of the dynamical variables (tensors) of CTM.\footnote{
CTM has a canonical conjugate pair of real symmetric three-index tensors, $Q_{abc}$ and $P_{abc}$ ($a,b,c=1,2,\ldots,N)$.
See \ref{app:ctm} for more about CTM.}
This peak-Lie group relation implies that Lie group invariant configurations are favored, or in other words, 
Lie group symmetries emerge in CTM.
Then, from the perspective mentioned above in the first paragraph, the question is whether  
this can be linked to emergence of spacetimes. 
In this paper, we show that this can be seen in $\Psi(Q)$, which is the wave function representing 
the same state in the other pair $Q$ conjugate to $P$.

The most important difference of this paper from the previous similar study \cite{Sasakura:2021lub} 
of $\Psi(Q)$ is the discovery of 
a new phase, which we call the classical phase in this paper. 
In this phase, fluctuations of variables are suppressed, and there emerge configurations which are
discretizations of classical geometric spaces.
This will explicitly be demonstrated for $n$-dimensional spheres $S^n\ (n=1,2,3)$ 
by computing the wave function $\Psi(Q)$ for $Q$ taking $SO(n+1)$ invariant or nearby values
by applying the Hamiltonian Monte Carlo method.

As we will see later, the transition to the classical phase has striking resemblance to a matrix counterpart, which is the transition 
from one-cut to two-cut solutions in the large $N$ limit of the matrix model \cite{Eynard:2016yaa}, 
or the Gross-Witten-Wadia type transition  \cite{Gross:1980he,Wadia:1980cp}.
In fact, in the previous paper \cite{Sasakura:2020rqz}, 
a two-logarithm matrix model which is a matrix analogue to the wave function of CTM was analyzed, and it was
shown that there exist transitions from one-cut to two-cut solutions where certain dimensional cloud-like configurations emerge.
Though these configurations of certain dimensions are hard to be regarded as spaces with classical geometry, 
it is interesting that we find a parallel phenomenon in the matrix model.   

As will be discussed later, we want to stress the importance of the positivity of the cosmological constant of CTM taken in this paper. 
When it is positive, the wavefunction is expressed by an oscillatory integration, that generally suppresses
the configurations in the quantum phase compared to those in the classical phase.  
Therefore the positivity is essentially important for the emergence of classical geometric spaces which appear 
in the classical phase.

This paper is organized as follows. 
In Section~\ref{sec:wavefn}, we explain the setup, namely, the wavefunction
of CTM we analyze.
In Section~\ref{sec:mc}, we explain the Monte Carlo method we employ, namely, the 
reweighting method applied to the wavefunction.
In Section~\ref{sec:symq}, we explain the way to construct the Lie group invariant values of $Q$ we take as the argument of the 
wavefunction. We consider $SO(n+1)\ (n=1,2,3)$ as the Lie groups, and take natural sets of representations on $Q$.
In Section~\ref{sec:twophases}, we show the presence of two phases, the quantum and the classical phases,
by the Monte Carlo method. We observe the transition that the topology of the distributions of the variables 
continuously changes between one bunch in the quantum phase and two bunches in the classical phase.  
In the classical phase, the fluctuations of the variables are suppressed.
In Section~\ref{sec:trd}, we discuss the translation between geometry and a real symmetric three-index tensor
through the tensor rank decomposition. In particular, we define a discrete analogue of Laplacian, which is used to analyze the
geometry of emerging configurations in the classical phase.     
In Section~\ref{sec:geometry}, we study the topology and the geometry of the emerging configurations in the classical 
phase. We find $S^{n}$ for $SO(n+1)\ (n=1,2,3)$ invariant $Q$.
In Section~\ref{sec:deform}, we consider two kinds of deformations of $Q$ from those given in Section~\ref{sec:symq}.
One is to change the representations on $Q$ and the other is to break the Lie group invariance. In both cases,
we find the classical phase becomes less likely by the deformations.
In Section~\ref{sec:oscillatory}, we study the behavior of the complex part in the reweighting method in some details. 
Because of the positivity of the cosmological constant, 
the configurations in the quantum phase are generally suppressed compared to those in the classical phase.
This suppression is enhanced for larger $|Q|$, and  
the main physical statement is that the system starts from the quantum phase with small $|Q|$, 
and undergoes the transition to the 
classical phase, as $|Q|$ develops, which is expected to be correlated with time. 
The last section is devoted to a summary and future prospects. 

\section{The wave function} 
\label{sec:wavefn}
In this section, we will explain the setup of the wave function we will analyze, leaving some details to \ref{app:ctm}.

Canonical tensor model (CTM) \cite{Sasakura:2011sq,Sasakura:2012fb} is a tensor model in the canonical formalism,  
formulated in the analogue of the ADM formalism \cite{Arnowitt:1962hi} of general relativity. 
The motivation for the formulation is to incorporate a time direction consistently with the fundamental of general
relativity, namely, the general covariance. 
The quantized Hamiltonian \cite{Sasakura:2013wza}
of CTM is given by
\[
\hat H=N_a {\hat H}_a +N_{[ab]} {\hat J}_{[ab]}, 
\] 
where ${\hat H}_a$ and ${\hat J}_{[ab]}$ are the  quantized Hamiltonian and momentum constraints of CTM, 
and $N_a$ and $N_{[ab]}$ are the lapse and the shift parameters, respectively. 
Here contracted lower indices are assumed to be summed over throughout this paper.
The commutation algebra between the quantized constraints is non-linearly closed, making them first-class constraints. 
The physical state condition is given by the CTM analogue of the Wheeler-DeWitt equation \cite{DeWitt:1967yk},
\[
{\hat H}_a |\Psi \rangle=0,
\label{eq:wdw}
\]
accompanied with $\hat J_{[ab]}|\Psi \rangle=0$.
An explicit solution for general $N$ exists \cite{Narain:2014cya} to these equations, 
and the wave function in the $P$-representation is given by an integral expression, 
\[
\Psi(P):=\langle P | \Psi \rangle= \int_{\cal C} \mathd\phi\, \mathd\tilde \phi\, \mathe^{i\sum_{j=1}^R \left(P(\phi^j)^3-(\phi^j)^2 \tilde \phi^j +(\tilde \phi^j)^3/3 \right)},
\label{eq:psip}
\]
where the integration variables are $\phi_a^j,\tilde \phi^j\ (a=1,2,\ldots,N,\ j=1,2,\ldots,R)$, and 
\s[
&\mathd\phi\, \mathd\tilde \phi:=\prod_{j=1}^R  \mathd\tilde \phi^j  \prod_{a=1}^N  \mathd\phi_a^j, \\
&P(\phi^j)^3:=P_{abc}\phi^j_a\phi^j_b \phi^j_c,   \\
&(\phi^j)^2:=\phi_a^j \phi_a^j.
\s]
Here summations over the upper indices (namely, $j$ above) will explicitly be indicated throughout this paper.
The parameter $R$ is restricted to be $R=(N+2)(N+3)/4$, which comes from the hermiticity of the Hamiltonian constraint ${\hat H}_a$
\cite{Sasakura:2013wza}. 
In the above expression \eq{eq:psip}, $R$ severs as the replica number of the set $(\phi_a^j,\tilde \phi^j)$, and 
the possible values of $N$ are restricted for $R$ to be an integer. 
However, we would be able to assume that small deviations of $R$ would not largely change the dynamics of our 
Monte Carlo simulations, and will take 
\[
R=\lfloor (N+2)(N+3)/4 \rfloor,
\label{eq:rval}
\]
where $\lfloor \cdot \rfloor$ denotes the floor function, allowing any values of $N$.
We are taking a positive value for the cosmological constant of CTM, since the positivity is essentially important 
for the peak-Lie group symmetry relation \cite{Obster:2017dhx,Obster:2017pdq} mentioned in Section~\ref{sec:introduction}. 
Correspondingly, the wave function \eq{eq:psip} is the expression for a positive cosmological constant of CTM \cite{Sasakura:2014gia}.
In the positive case, the second term in the exponent of \eq{eq:psip} takes the minus sign as shown there,
and this sign plays some essential roles in later sections.
As for the integration region ${\cal C}$, we take the integration region of $\phi$ to be  
the real numbers ${\mathbb R}^{NR}$,\footnote{
The integration region must be deformed slightly from the real values to make the integration 
convergent, or a regularization must be introduced as in \cite{Obster:2017dhx}. 
A mathematically rigorous way to define the integration contour can be provided by the Lefschetz thimble \cite{Witten:2010cx}.}
while the integration contour of $\tilde \phi$ will be specified in Section~\ref{sec:mc}.   

The wave function in $Q$-representation can be obtained by the Fourier transformation of \eq{eq:psip}. However, this generates a product
of delta functions because the exponent of the integrand is linear in $P$. This delta function product is difficult to handle in 
the Monte Carlo method. 
Therefore we introduce a regularization term $-P_{abc}P_{abc}/4 \lambda\ (\lambda>0)$ into the exponent, and then we obtain
\s[
\Psi(Q,\lambda)&=\int  \mathd P  \mathd\phi\,  \mathd\tilde \phi\, \mathe^{-i P_{abc}Q_{abc} -P_{abc}P_{abc}/4\lambda}\, \Psi(P)  \\
&=\int_{\cal C}  \mathd\phi\,  \mathd\tilde \phi \,
\mathe^{-\lambda \left(Q-\sum_{j=1}^R \phi^j\phi^j\phi^j\right)^2+i\sum_{j=1}^R \left(-(\phi^j)^2 \tilde \phi^j +(\tilde \phi^j)^3/3 \right)},
\label{eq:psiq}
\s]
where we have ignored an inessential overall factor, and have introduced a short-hand notation,
\[
\left(Q-\sum_{j=1}^R \phi^j\phi^j\phi^j\right)^2:=\left(Q_{abc}-\sum_{j=1}^R \phi_a^j\phi_b^j\phi_c^j\right)
\left(Q_{abc}-\sum_{j=1}^R \phi_a^j\phi_b^j\phi_c^j\right).
\]

In the above, the parameter $\lambda$ is introduced as a regularization parameter, which changes 
the wave function. Since the wave function should be the solution to the Wheeler-DeWitt equation \eq{eq:wdw}, 
the physical meaning of $\Psi(Q,\lambda)$ with a finite $\lambda$ is not well given.
Instead, rather than as a regularization, we may introduce $\lambda$ as a part of an observable which
operates on $\Psi(q):=\int  \mathd P \mathe^{-i\, P_{abc} q_{abc}}\, \Psi(P)$, which is the genuine wave function in $Q$-representation.
More precisely, it is introduced as a smearing operator $\langle q_1 | \hat O_\lambda | q_2 \rangle:=\mathe^{-\lambda (q_1-q_2)^2}$ 
acting on $\Psi$:
\[
\Psi(Q,\lambda):=(\hat O_\lambda \Psi )(Q)=\int  \mathd q \ \mathe^{-\lambda(Q-q)^2} \Psi(q),
\] 
which has the same form as \eq{eq:psiq} up to an irrelevant normalization.

In the former way of introducing $\lambda$ above, it must be taken infinitely large to remove the regularization. In the latter 
way above, it is not necessary to take it infinitely large, but,  
as we will see later, the coupling $\lambda$ is effectively replaced by $\lambda |Q|^2$. This means that
the coupling must be taken larger as spacetime develops.\footnote{Here we assume
$|Q|$ is correlated with time, as  $Q$ is proportional to the spatial volume in the agreement of the $N=1$ CTM 
with the mini-superspace approximation of general relativity \cite{Sasakura:2014gia}.\label{foot:q}}
However, we consider only the range $\lambda \lesssim 10^7$ in this paper,
which comes from the technical reason that our Monte Carlo simulation becomes inefficient above this value.

It is worth commenting on the convergence of \eq{eq:psiq}. The sum $\sum_{i=1}^R \phi_a^i\phi_b^i\phi_c^i$ contained in 
\eq{eq:psiq} has a lot of flat directions to the infinity of $\phi$, 
such as the one with $\phi^i_a \phi^i_b \phi^i_c+\phi^j_a \phi^j_b \phi^j_c=0\ (i\neq j)$  by 
taking $\phi^i=-\phi^j$. Therefore it is generally a non-trivial matter whether the integral \eq{eq:psiq} converges or not.
The convergence for $R\lesssim N^2/2$ was first noticed in \cite{Obster:2020vfo}, and was systematically analyzed 
in \cite{Obster:2021xtb}. Since our value of $R$ in \eq{eq:rval} is roughly smaller by a factor of 2, the current analysis 
does not suffer from the divergence, which indeed was checked in our actual Monte Carlo simulations. 

\section{The Monte Carlo method}
\label{sec:mc}
We want to evaluate \eq{eq:psiq} by the Monte Carlo method. However, 
the integrand contains an oscillatory part, namely, \eq{eq:psiq} suffers from the notorious sign problem \cite{Berger:2019odf}. 
To deal with this issue we take the most naive method,  the so-called reweighting method.
Though more sophisticated methods exist, it would be appropriate to apply this simple method to the current primary study 
of the system, since it would presently be difficult to foresee potential complications caused by the other more sophisticated methods.

\begin{figure}
\begin{center}
\includegraphics[width=6cm]{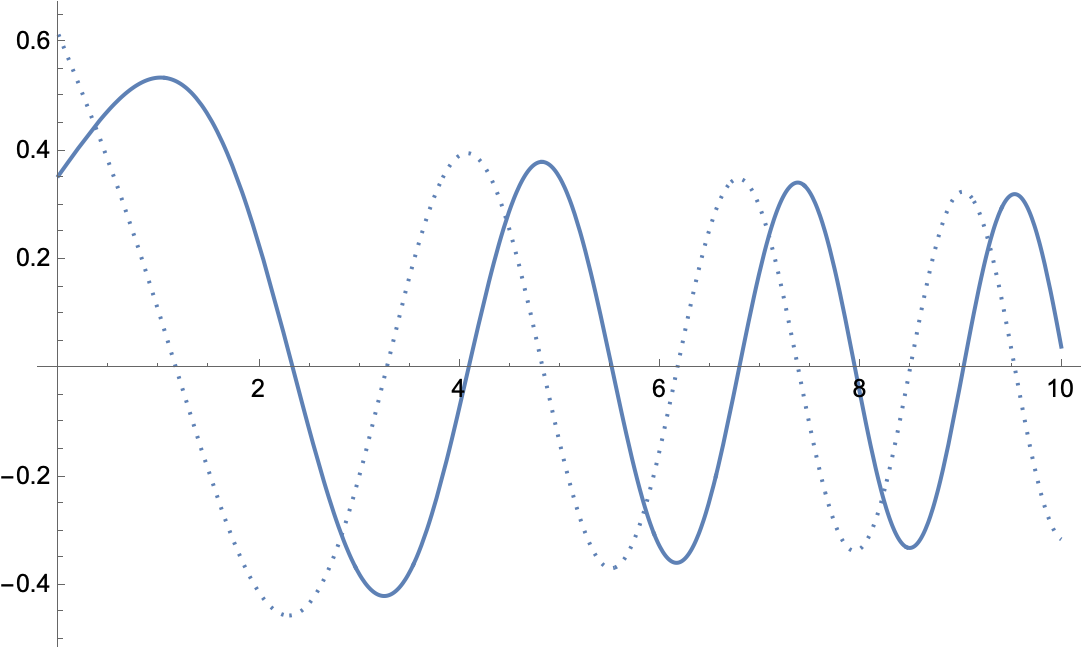}
\hfil
\includegraphics[width=4cm]{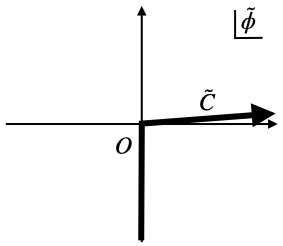}
\caption{Left: The plot of \eq{eq:actairy} against $z$. Solid and dashed lines represent the real and imaginary parts, respectively.
Right: Integration contour $\tilde C$, slightly deformed from the real axis.}
\label{fig:airy}
\end{center}
\end{figure}

Let us first rewrite 
\[
\Psi(Q,\lambda)=Z_{Q,\lambda} \left \langle \int_{\tilde C}  \mathd\tilde \phi\,
\mathe^{i\sum_{j=1}^R \left(-(\phi^j)^2 \tilde \phi^j +(\tilde \phi^j)^3/3\right) }
\right\rangle_{Q,\lambda},
\]
where
\[
&Z_{Q,\lambda}=\int_{\mathbb{R}^{NR}}  \mathd\phi\,  \mathe^{-\lambda \left(Q-\sum_{j=1}^R \phi^j\phi^j\phi^j\right)^2},
\label{eq:defofz}
\]
and the expectation value $\langle \cdot \rangle$ is taken in the system defined by the partition function \eq{eq:defofz}.
The expression becomes more convenient for actual Monte Carlo simulations by splitting the size of $Q$ as
$Q_{abc}=|Q| \tilde Q_{abc}$, where $|Q|:=\sqrt{Q_{abc}Q_{abc}}$ and performing the rescaling $\phi\rightarrow |Q|^{1/3} \phi$:
\[
 \Psi(Q,\lambda)=|Q|^{NR/3} \, Z_{\tilde Q, \lambda |Q|^2}\,
 \left \langle 
 \prod_{j=1}^R {\rm Airy}\left(-|Q|^{2/3} (\phi^j)^2\right) \right\rangle_{\tilde Q,\lambda |Q|^2},
 \label{eq:repsiq}
\]
where we have used the integral expression of the Airy function,
\[
{\rm Airy}(-z)=\int_{\tilde C}   \mathd\tilde \phi \, \mathe^{i \, \left(-z \tilde \phi + \tilde \phi^3/3\right)}.
\label{eq:airy}
\]
Note that our interest is only the positive region $z\geq 0$ as in \eq{eq:repsiq}.
A natural choice of the Airy function for the current study is given by setting 
\[
{\rm Airy}(-z)={\rm Ai}(-z)+i \, {\rm Bi}(-z), 
\label{eq:actairy}
\]
since this function asymptotically approaches a plane wave 
form (See Figure~\ref{fig:airy}).  This choice corresponds to the integration contour $\tilde C$ to be as in the 
right panel of Figure~\ref{fig:airy}.
From \eq{eq:repsiq}, we see that it is enough to compute $Z_{Q,\lambda}$ and 
$\langle \cdot \rangle_{Q,\lambda}$  for $|Q|=1$, which we will carry out in later sections.

We employ the Hamiltonian Monte Carlo method \cite{Neal(2011)} 
to generate the sampling sequence of $\phi_a^j\, (a=1,2,\ldots,N,\ j=1,2,\ldots,R)$
for the system defined by \eq{eq:defofz}. The leapfrog numbers are typically taken with a few hundreds
depending on the sizes of $N,R$. We also use parallel tempering \cite{paralleltemp} across different values of $\lambda$
for a difficult case (More concretely, the $SO(4)$ case with $N=30$ and $R=264$, which will appear later.).
The total numbers of the samples for each sequence are typically around $10^4\sim 10^6$.
As for the analyzed data, one data is taken from every $\sim 10^2$ samples to remove correlations.
The machine has a Xeon W2295 (3.0GHz, 18 cores), 128GB DDR4 memory, and Ubuntu 20 as OS.  
The program is written in C++ with the use of pthread for parallelization. As for the Airy function, the boost 
library \cite{boost} is used. Every run typically takes several hours with active use of parallelization.

Lastly, we want to stress the importance of taking the cosmological constant of CTM to be positive in our setup.
In this case, the Airy function is oscillatory, as is taken above, 
and this oscillatory property plays the very important role of highlighting the configurations in the classical phase,
as will be discussed in Section~\ref{sec:oscillatory}. 
On the other hand,  if the cosmological constant is taken negative, the first term in the exponent of \eq{eq:airy} takes the positive sign. 
In this case, $\hbox{Airy}(z)$ is a linear combination of the two Airy functions,  $\hbox{Ai}(z)$ and $\hbox{Bi}(z)$, which are asymptotically 
exponential\footnote{They behave $\hbox{Ai}(z)\sim \mathe^{-2 z^{3/2}/3}$ 
and $\hbox{Bi}(z)\sim \mathe^{2 z^{3/2}/3}$  for $z\rightarrow +\infty$.},
but the only possible choice will be $\hbox{Airy}(z)=\hbox{Ai}(z)$ to avoid divergence in $z\rightarrow \infty$.
Now $\hbox{Ai}(z)$ is a monotonically damping function, and does  not
discriminate the configurations in the classical phase from those in the quantum phase. 
In fact, in \cite{Lionni:2019rty,Sasakura:2019hql,Obster:2020vfo},
the wavefunction in the negative cosmological constant case was analyzed with an approximation
to the Airy function part, but no signs of emerging classical geometric spaces have been found. 

\section{Construction of Lie group invariant $Q$}
\label{sec:symq}
There exist various manners to construct Lie group invariant $Q$. 
We employ the construction using harmonic functions on Lie group invariant manifolds \cite{Kawano:2018pip},
with the expectation for the emergence of space-like objects from the wave function \eq{eq:psiq} for such $Q$. 
Some deformations from such $Q$ will also be considered in Section~\ref{sec:deform}.

In this paper, we consider three Lie groups, $SO(n+1)\ (n=1,2,3)$.
The manifolds we expect to emerge are $S^n$, respectively.

\subsection{$SO(2)$ invariant $Q$}
The harmonic functions on $S^1$ are given by
\[
\{ f_a(\theta) \}=\{1/\sqrt{2}\} \cup \{ \cos(p\, \theta),\sin(p\,\theta)\, |\, p=1,2,\ldots,\Lambda \},
\]
where $\Lambda$ denotes a cut-off of the momentum. There are in total $N=2\Lambda+1$ functions. 
Then an invariant tensor $Q$ is constructed by
\[
Q^{SO(2)}_{abc}(\alpha)=const. \mathe^{-\alpha (p_a^2+p_b^2+p_c^2)/\Lambda^2 }\int_{0}^{2 \pi}  \mathd\theta \, f_a(\theta) f_b(\theta) f_c(\theta),
\]
where $p_a$ denotes the momentum of the function $f_a$ (namely, $p_a$ for  $f_a=\cos(p_a\theta),\sin(p_a \theta)$).  
$const.$ is a normalization factor for $|Q^{SO(2)}(\alpha)|=1$. The exponential factor introduces a smooth momentum cut-off,
and larger $\alpha$ would effectively make $S^1$ smaller, since the number of functions is effectively reduced.     
A negative $\alpha$ would also be possible but a space represented by such $\alpha$ would be pathological. 

\subsection{$SO(3)$ invariant $Q$}
The harmonic functions on $S^2$ are the spherical harmonics $Y_{l}^{m}(\omega)$, where $\omega$ is a coordinate system on 
$S^2$. Similarly, we consider 
\[
\{f_a\}=\{ {\rm Re} \,Y_l^m(\omega), {\rm Im}\, Y_l^m(\omega) | l=0,1,\ldots, \Lambda,\, m=-l,-l+1,\ldots,l\},
\]
where vanishing functions and degenerate ones are supposed to be ignored (ex., $\hbox{Im }Y_{l}^0=0$).  
There are in total $N=(\Lambda+1)^2$ independent functions.
Then we define\footnote{The $l$ dependence of the regularization term can be taken differently like using the eigenvalues of the laplacian 
instead. But such details would not affect the essential results as far as the regularization is a smooth damping function invariant 
under the Lie group. Otherwise the results cannot be considered universal.}  
\[
Q^{SO(3)}_{abc}(\alpha)=const.\, \mathe^{-\alpha (l_a^2+l_b^2+l_c^2)/\Lambda^2 }\int_{S^2} \mathd^2 \omega 
\, f_a(\omega) f_b(\omega) f_c(\omega),
\]
where $const.$ is a normalization factor for $|Q^{SO(3)}(\alpha)|=1$.

\subsection{$SO(d+1)$ invariant $Q$}

The harmonic functions on $S^d$ are $Y_{m,l_1,\ldots,l_{d-1}}(\omega)$, where $(\omega^i)=(\varphi,\theta_1,\ldots,\theta_{d-1})$ 
is a spherical coordinate system on $S^d$. Similarly, we consider
\begin{align}
\{f_a\}&=\{Y_{m,l_1,\ldots,l_{d-1}}|m\in\mathbb{Z},l_i\in\mathbb{Z}_{\geq 0},|m|\leq l_1\leq l_2\leq\cdots\leq l_{d-1}\leq\Lambda\}.
\end{align}
An iterative way of constructing $Y_{m,l_1,\ldots,l_{d-1}}(\omega)$ is given as follows 
\cite{gegenbauer}: given the harmonic functions $Y_{m,l_1,\ldots,l_{d-2}}
(\varphi,\theta_1,\ldots,\theta_{d-2})$ on $S^{d-1}$, the harmonic
functions on $S^d$ is defined by
\begin{align}
Y_{m,l_1,\ldots,l_{d-1}}(\varphi,\theta_1,\ldots,\theta_{d-1})&=B_{l_{d-1},l_{d-2}}^{d}(\theta_{d-1}) \, Y_{m, l_1, \ldots, l_{d - 2}}(\varphi,\theta_1,\ldots,\theta_{d-2}),
\label{eq:genY}
\end{align}
where
\[
B_{l_{d-1},l_{d-2}}^d(\theta): =\left( 
\frac{\left(l_{d-1}+\frac{d-1}{2}\right)(l_{d-1}+l_{d-2}+d-2)!}{(l_{d-1}-l_{d-2})!} \right)^{\frac{1}{2}}(\sin\theta)^{1-\frac{d}{2}}P_{l_{d-1}-1+\frac{d}{2}}^{-l_{d-2}+1-\frac{d}{2}}(\cos \theta),
\]
with the Legendre function $P$.
The derivation is summarized in \ref{app:spherical}.
Since harmonic functions satisfy the Helmholtz equation,
\begin{align}
-\Delta Y_{m,l_1,\ldots,l_{d-1}}&=l_{d-1}(l_{d-1}+d-1)Y_{m,l_1,\ldots,l_{d-1}}
\end{align}
with the Laplacian $\Delta$, the eigenvalue associated with $f_a=Y_{m,l_1,\ldots,l_{d-1}}$ is $\lambda_a=l_{d-1}(l_{d-1}+d-1)$. The total number of independent functions is $N=[\Lambda^{\bar{d}}+(\Lambda+1)^{\bar{d}}]/d!$, where $x^{\bar{n}}=x(x+1)\cdots(x+n-1)$ is the rising factorial. 

Now let us look at the case where $d=3$. Since $Y_l^m$ in the standard notation corresponds to $Y_{m,l}$ in the case of $d=2$, the subset of harmonic functions on $S^3$ is
\begin{align}
\{f_a\}&=\{B_{k,l}^3(\chi) \,Y_l^m(\varphi,\theta)|m\in\mathbb{Z},l,k\in\mathbb{Z}_{\geq 0},|m|\leq l\leq k\leq\Lambda\}.
\end{align}
The cardinality of the set is $N=(\Lambda+1)(\Lambda+2)(2\Lambda+3)/6$. Then we define
\begin{align}
Q_{abc}^{SO(4)}(\alpha)&=const.\mathe^{-\alpha(\lambda_a+\lambda_b+\lambda_c)/\lambda_\Lambda}\int_{S^3}
\mathd^3\omega\, f_a (\omega)f_b(\omega)f_c(\omega),
\end{align}
where $const.$ is a normalization factor for $|Q^{SO(4)}(\alpha)|=1$.
To do this calculation, we used the Mathematica package HFT11 \cite{HFT}.
This package can enumerate a system of orthogonal functions on $S^d$ as a function of unit vector $\hat x\in\mathbb R^{d+1}$, and can perform integration over $S^d$. 

\section{Presence of two phases}
\label{sec:twophases}
\begin{figure}
\begin{center} 
\includegraphics[width=3cm]{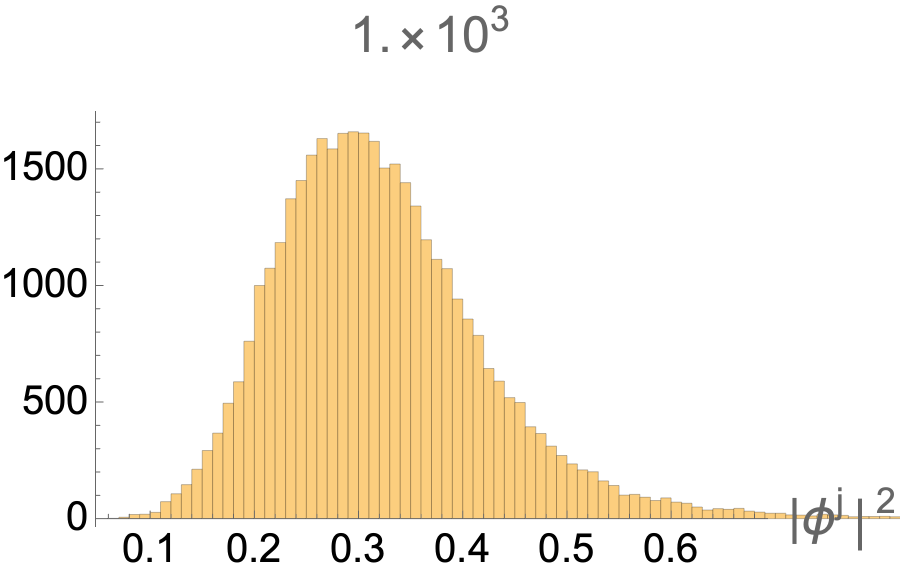}
\hfil
\includegraphics[width=3cm]{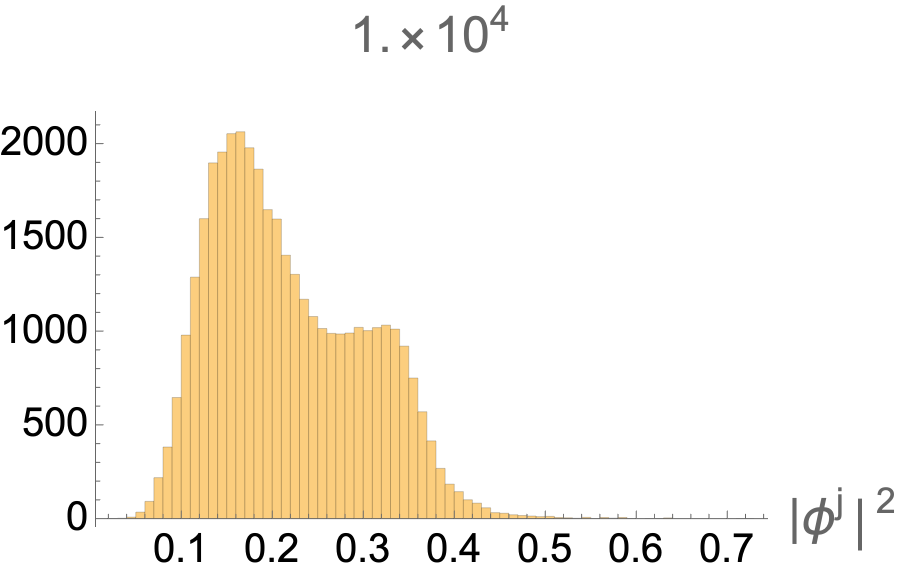} 
\hfil
\includegraphics[width=3cm]{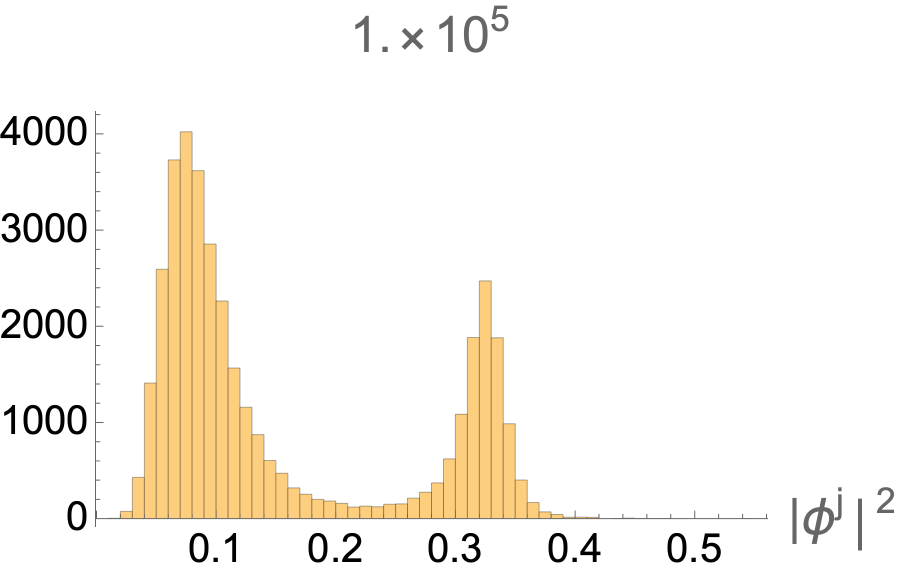}
\hfil
\includegraphics[width=3cm]{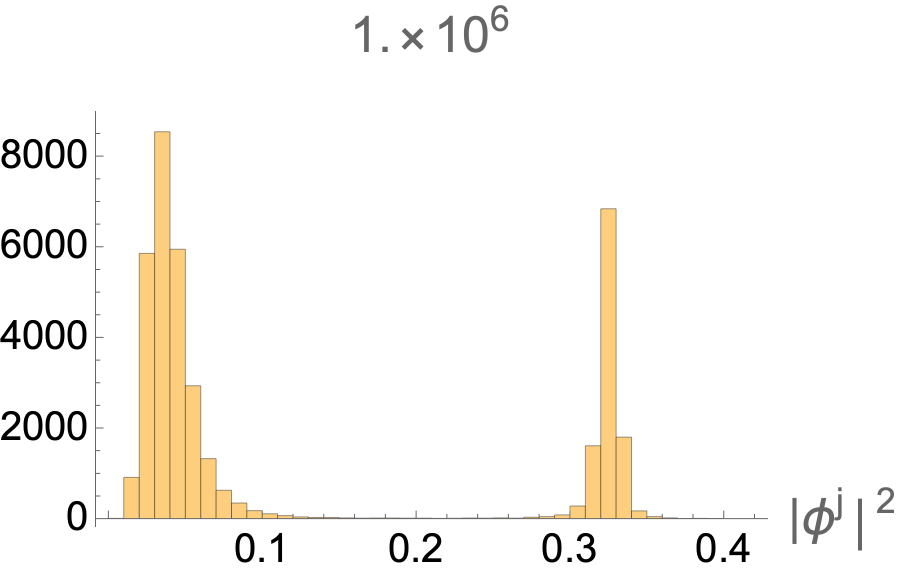}
\caption{The histograms of $|\phi^j|^2$ for $Q^{SO(2)}$ with $N=15\ (\Lambda=7)$ and $\alpha=0.5$. 
The data are the collection of $|\phi^j|^2\ (j=1,2,\ldots,R)$ over the sampled data of the Monte Carlo simulations. 
The values of $\lambda$ are shown over each figure.
The classical phase can be found for $\lambda\gtrsim10^6$.}
\label{fig:so2tran}
\end{center}
\end{figure}

\begin{figure}
\begin{center}
\includegraphics[width=3cm]{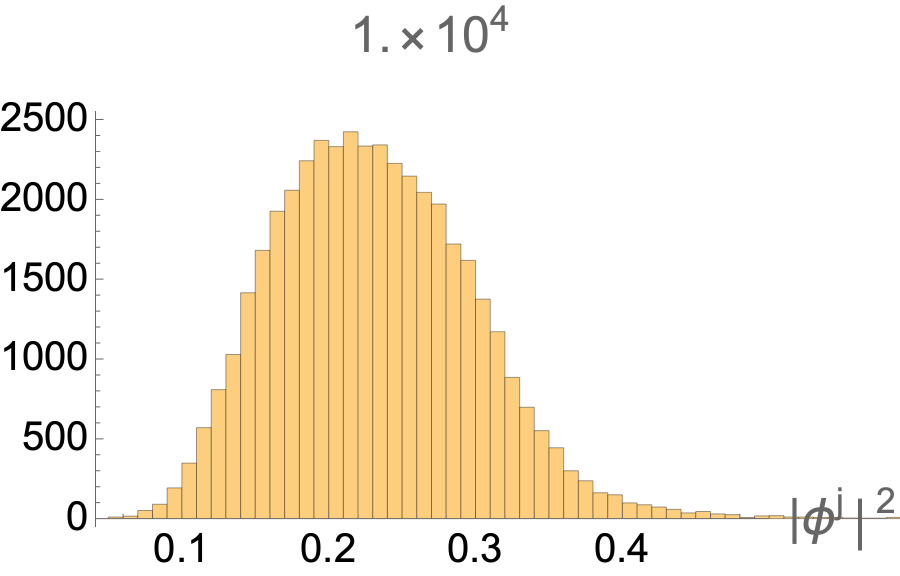}
\hfil
\includegraphics[width=3cm]{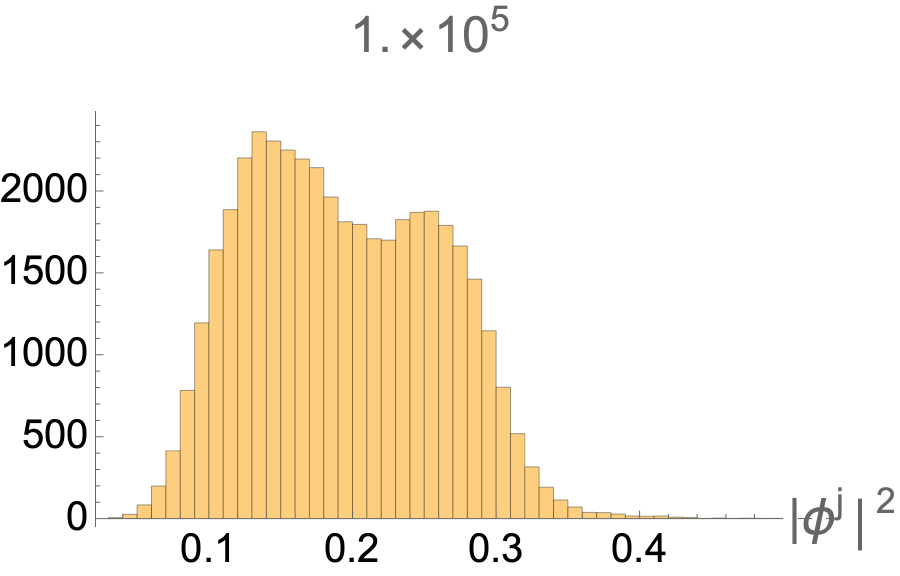} 
\hfil
\includegraphics[width=3cm]{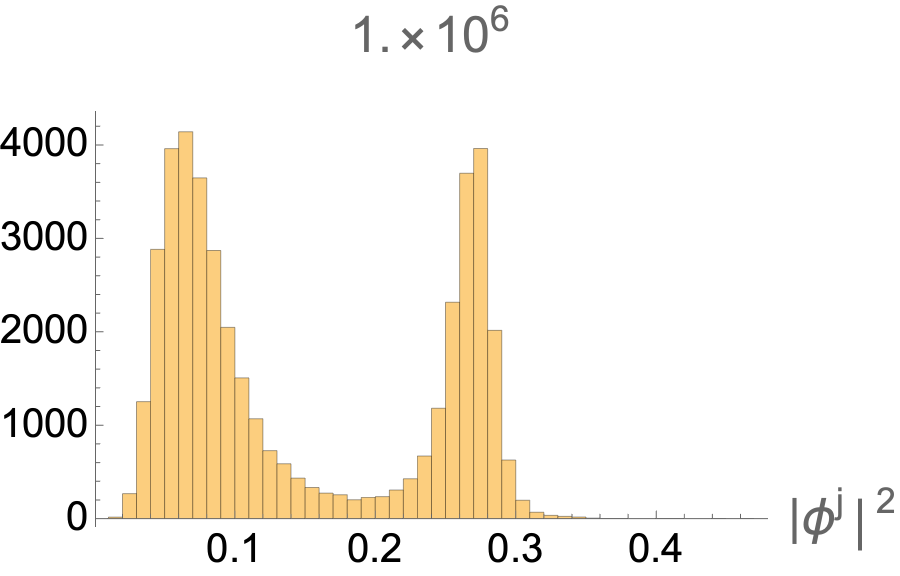}
\hfil
\includegraphics[width=3cm]{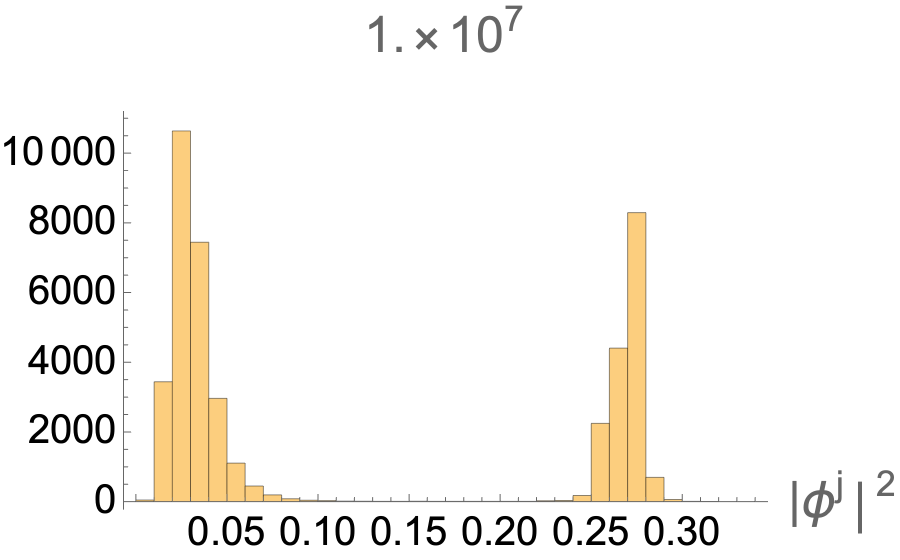}
\caption{The histograms for $Q^{SO(3)}$ with $N=16\ (\Lambda=3)$ and $\alpha=0.5$, 
plotted in the same way as in Figure~\ref{fig:so2tran}.
The classical phase can be found for $\lambda\gtrsim10^7$.}
\label{fig:so3tran}
\end{center}
\end{figure}

\begin{figure}
\begin{center}
\includegraphics[width=3cm]{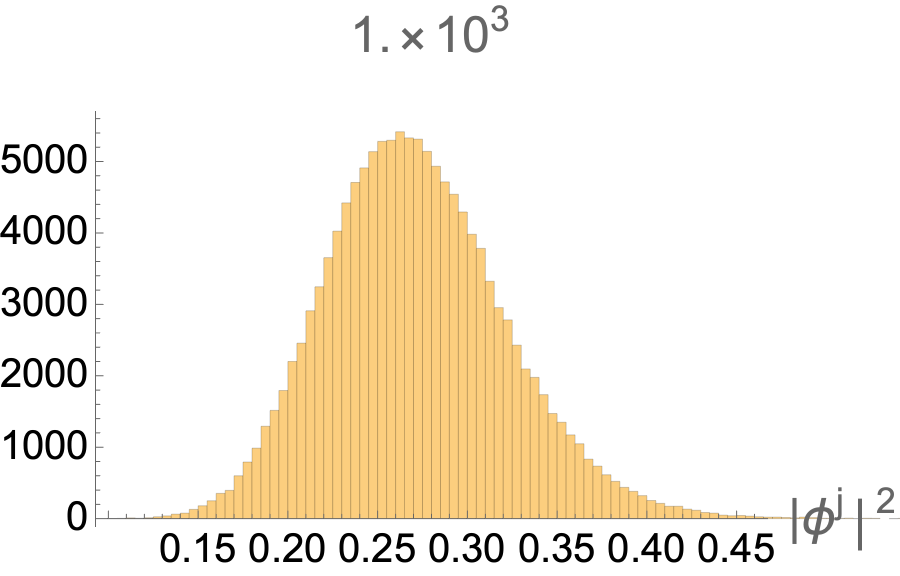}
\hfil
\includegraphics[width=3cm]{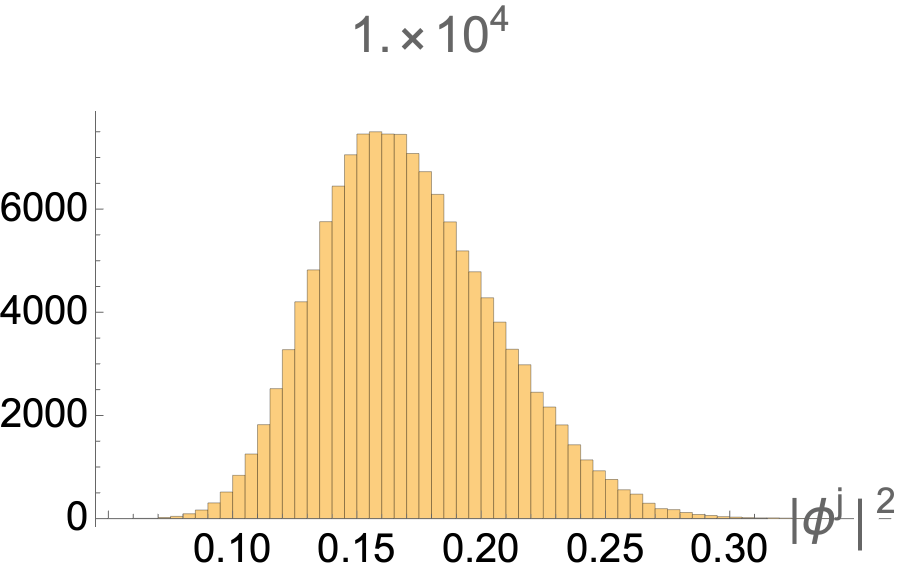} 
\hfil
\includegraphics[width=3cm]{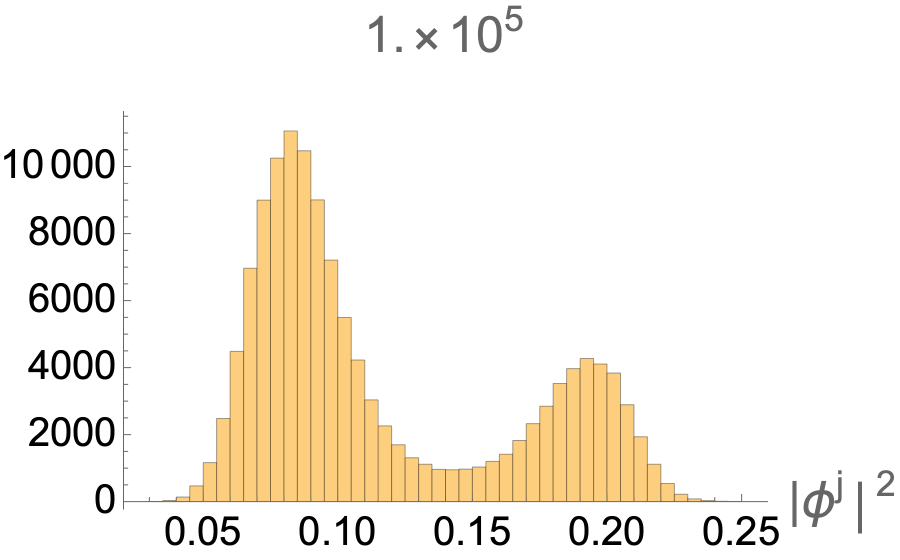}
\hfil
\includegraphics[width=3cm]{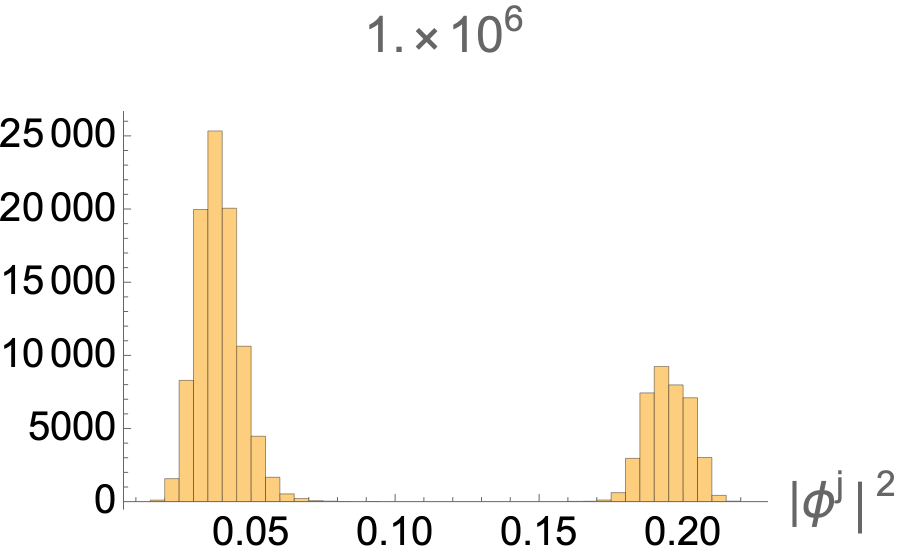}
\caption{The histograms for $Q^{SO(4)}$ with $N=30\ (\Lambda=3)$ and $\alpha=0.5$, 
plotted in the same way as in Figure~\ref{fig:so2tran}.
The classical phase can be found for $\lambda\gtrsim10^6$.
On the other hand, $\Lambda=2$ case does not show the presence of the classical phase up to $\lambda\leq 10^7$. }
\label{fig:so4tran}
\end{center}
\end{figure}

In this section, we will show that, for the Lie-group invariant $Q$'s given in Section~\ref{sec:symq}, the system defined by 
the partition function \eq{eq:defofz} has two phases characterized by the topology of the distribution of $|\phi^j|^2$.
The histograms of the distributions for $Q^{SO(2)}$, $Q^{SO(3)}$, and $Q^{SO(4)}$ are shown in Figure~\ref{fig:so2tran}, 
Figure~\ref{fig:so3tran}, and Figure~\ref{fig:so4tran}, respectively.
Here we often take $\alpha=0.5$ in this paper, because the classical phase is most evident 
in the range $0\lesssim \alpha \lesssim 1$, as will be explained more in Section~\ref{sec:oscillatory}.
As can been seen in the histograms, the two phases can be characterized by the number of the connected regions of the distributions,
one for smaller $\lambda$ and two for larger $\lambda$. 
An important property of the latter phase
is that the widths of each bunch of the distributions are smaller than that in the former. 
Because of this suppression of the fluctuations, 
we call the former and the latter phases the quantum and the classical phases, respectively, as we will see more evidences of 
classicality of the latter phase. 
We also call the two bunches of the distributions in the classical phase the central and outer bunches, respectively, 
based on the distances from the origin.


\begin{figure}
\begin{center}
\includegraphics[width=10cm]{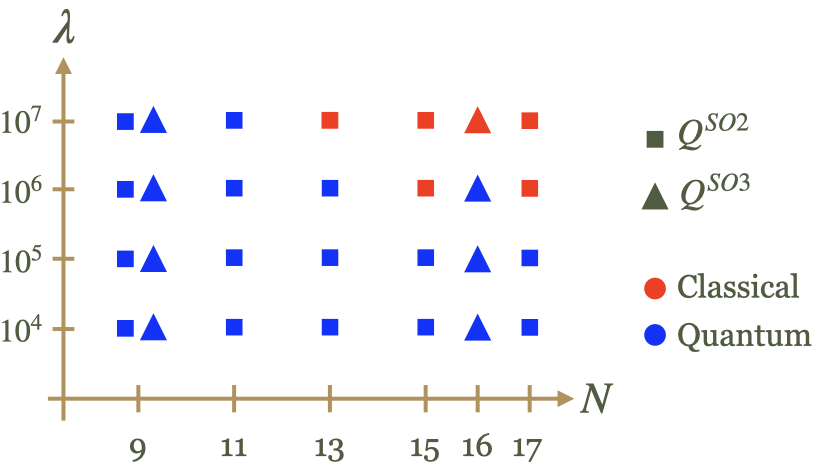}
\caption{The values of $\lambda,N$ are classified according to the phases for $Q^{SO(2)}$ and $Q^{SO(3)}$ with $\alpha=0.5$.
The classification here is qualitative and 
contains some ambiguities, since it is based on the appearance of histograms
but not by a quantitative measure. }
\label{fig:classreg}
\end{center}
\end{figure}

Figure~\ref{fig:classreg} classifies the values of $\lambda$ and $N$ according to the phases. 
In general the classical phase appears, when $\lambda$ and $N$ are large.  Note that this classification is merely qualitative,
as it is based merely on the appearance of the histograms, not by a quantitative measure. 
A possible quantitative measure could be given by the method developed in \cite{Hartnoll:2019pwe}.
The application of such measures to our setup is left for future study.

The transition resembles a matrix model counterpart, namely, the transition between one-cut and two-cut solutions 
of the matrix model in the large $N$ limit \cite{Eynard:2016yaa}, or the Gross-Wadia-Witten type transition \cite{Gross:1980he,Wadia:1980cp}. However, there is 
a difference in the mechanism of the distribution in our setup from that in the matrix model. 
When $\lambda$ is large, the partition function \eq{eq:defofz} approximately imposes a relation, 
\[
Q_{abc}= \sum_{j=1}^R \phi_a^j \phi_b^j \phi_c^j.
\label{eq:trd}
\]
This is a decomposition of a tensor $Q$ into a number of vectors $\phi^j$, and is known as the tensor rank decomposition
(also often called CP decomposition) \cite{SAPM:SAPM192761164,Carroll1970} in applied mathematics \cite{Landsberg2012}.
It is used for various purposes to extract information from tensors 
constructed from real-life data \cite{comon:hal-00923279}. 
The tensor rank decomposition can also be used to extract geometric information from tensors \cite{Kawano:2018pip}.
The minimum value of $R$ which realizes the decomposition \eq{eq:trd} for a given $Q$ is called the rank\footnote{
More precisely, in the current case of real $Q,\phi$ and the symmetric form on the righthand side of \eq{eq:trd}, 
the rank is called the real symmetric rank of $Q$.} of $Q$.
In the cases of Figure~\ref{fig:so2tran} and Figure~\ref{fig:so3tran}, the ranks of $Q$'s 
are smaller than the values of  $R$ (namely, the value in \eq{eq:rval}).\footnote{
The $Q$'s of Figure~\ref{fig:so2tran} and Figure~\ref{fig:so3tran} can be decomposed into 22 and 32 $\phi^j$'s, respectively.
This can be shown numerically (up to machine precisions) by using the program used in \cite{Kawano:2018pip}.}
 This generally means that there exist abundance of solutions of $\phi$ to \eq{eq:trd}, and 
they appear as the distributions in the figures.
On the other hand,  in the matrix model, 
the eigenvalue distribution comes from the distribution of the matrix itself,
while the tensor $Q$ is an external parameter in our setup.
 
\begin{figure}
\begin{center}
\includegraphics[width=5cm]{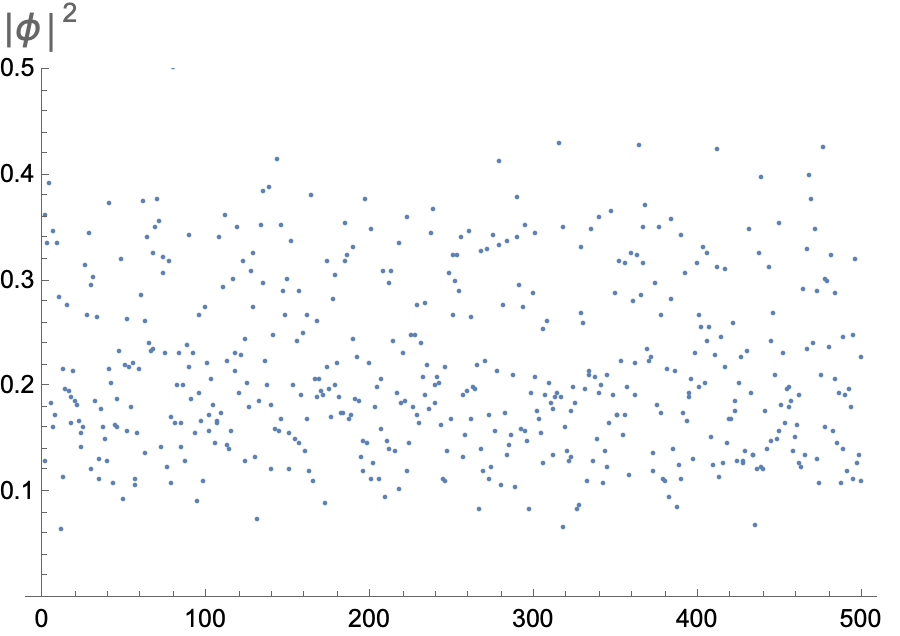}
\hfil
\includegraphics[width=5cm]{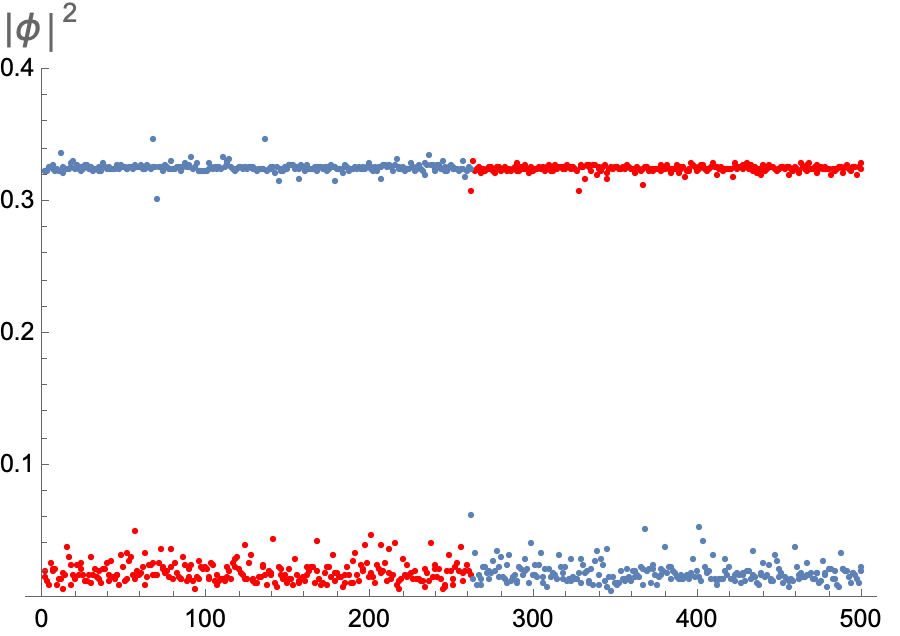}
\hfil
\includegraphics[width=5cm]{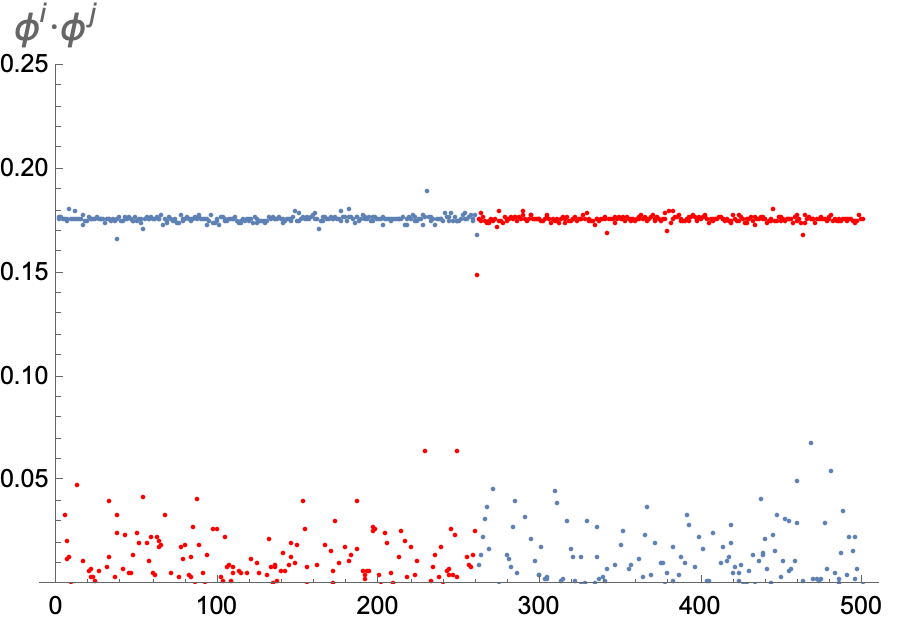}
\caption{The left panel: The MC sequence of one of the $|\phi^j|^2$'s for $Q^{SO(2)}$ with $N=15\ (\Lambda=7)$, $\lambda=10^4$,
and $\alpha=0.5$.
The vertical axis represents $|\phi^j|^2$ and the horizontal the sequence. The fluctuation of $|\phi^j|^2$ is wide. 
The middle panel: Two of $|\phi^j|^2$'s are plotted with different colors for $\lambda=10^7$; one in the central bunch and the other 
in the outer bunch.  The fluctuations around each center are strongly suppressed.
However there exists an exchange that one, say $\phi^1$, moves from the outer bunch to the center, while the other,
say $\phi^2$, moves in the opposite direction at the same time. 
They inherit each other's role.
The right panel: At the exchange $\phi^1\leftrightarrow \phi^2$, 
the inner products, $\phi^1_a \phi^3_a$ and $\phi^2_a \phi^3_a$,  are exchanged, where $\phi^3$ 
is in the outer bunch, but there are no net changes of the set of values.  }
\label{fig:mc}
\end{center}
\end{figure}

\begin{figure}
\begin{center}
\includegraphics[width=5cm]{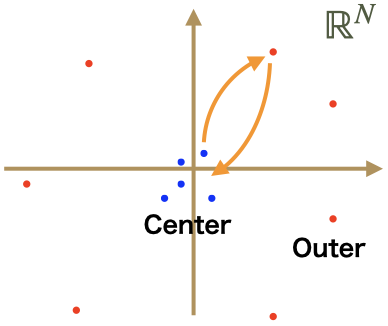}
\caption{The illustration of an exchange of $\phi^j$'s between the center and the outer bunches.
Such an exchange does not occur (or seem to be strongly suppressed) between $\phi^j$'s in the outer bunch.}
\label{fig:exchange}
\end{center}
\end{figure}

It is interesting to look at the actual Monte Carlo sequences. 
Figure~\ref{fig:mc} shows a few examples of the sequences of $|\phi^j|^2$ and $\phi^i_a \phi^j_a\ (i\neq j)$ in the two phases. 
The left panel corresponds to
the quantum phase, and the middle and right panels the classical. 
In the quantum phase,  $|\phi^j|^2$ fluctuates over a wider region.
On the other hand, in the classical phase, the fluctuations are confined around each center of the two bunches,
strongly suppressed in the examples (How strongly it is suppressed depends on the parameters.). 
Interestingly, as shown in the middle and the right panels, we still observe exchanges of $\phi^j$'s between the two bunches
in our Monte Carlo sequences: one $\phi^j$ in the center bunch moves into the outer bunch, and at the same time one in the outer bunch
moves into the center bunch (See Figure~\ref{fig:exchange} for an illustration.). However, the exchange is just a replacement
of the roles: there are no net changes of the values of $|\phi^j|^2$ and $\phi^i_a \phi^j_a$ in the outer bunch. 
Such an exchange does not seem to occur (or is strongly suppressed) between two $\phi^j$'s in the outer bunch.  
 
This observation suggests the following picture about the classical phase. Suppose that it has been confirmed that the distributions,
namely the center and the outer bunches, are completely separated in a certain thermodynamic limit, 
as in the two-cut solutions of the matrix model in the large $N$ limit.
Then 
\[
\{ \phi^i_a \phi^j_a \  |\  \forall\phi^{i},\phi^j\, \in \hbox{Outer bunch}\}
\label{eq:phi2outer}
\]  
provides a set of semi-classical order parameters characterizing the classical phase.
The inner products $\phi_a^i \phi_a^j$ are called overlaps and used for order parameters in the replica method 
in the spin-glass theory \cite{pedestrians}. The pattern of \eq{eq:phi2outer} generally breaks the replica symmetry, and 
the transition to the classical phase can be regarded as a replica symmetry breaking. 
In Section~\ref{sec:geometry} we will associate the pattern of this set of order parameters with classical geometric structure
of emergent spaces.

\section{Tensor rank decomposition and geometry}
\label{sec:trd}

\subsection{Integral representation of $Q$}
\label{subsec:intq}

In CTM, the dynamical variable $Q_{abc}$ is interpreted as a data set representing the geometry of a Cauchy hypersurface $\varSigma$. 
A specific method to obtain $Q$ corresponding to a given $d$-dimensional compact Riemannian manifold 
$(\varSigma,\gamma)$ was given in \cite{Kawano:2018pip}. 
What is done in Section~\ref{sec:symq} can be regarded as some special cases of the general method. 
The method is briefly summarized below.
\begin{enumerate}[1.]

\item Let $\alpha$ run from 1 to $d$, and let $(x^{\alpha})$ be the local coordinate on $\varSigma$. The Laplace-Beltrami operator $\Delta$ is defined by
\begin{align*}
\Delta&=\frac{1}{\sqrt{\gamma}}\partial_{\alpha}\left(\sqrt{\gamma}\gamma^{\alpha\beta}\partial_{\beta} \right)
\end{align*}
from the metric tensor $\gamma_{\alpha \beta}$ on $\varSigma$ and its determinant $\gamma=\det (\gamma_{\alpha \beta})$. 
 
\item Since $\varSigma$ is a compact manifold, $-\Delta$ on $\varSigma$ has a discrete spectrum, and we take $N$ of its eigenvalues, starting with the smallest, and index them so that $0 \leq \lambda_1 \leq \cdots \leq \lambda_N$ is satisfied. 
Furthermore, let $f_a$ be the eigenfunction belonging to the eigenvalue $\lambda_a$ and be orthonormalized like
\begin{align}
\delta_{ab}&=\int_{\varSigma}f_a (x)f_b (x)\sqrt{\gamma}\mathd^d x.
\label{eq:normf}
\end{align}
When $\varSigma$ is a closed manifold, or when it has boundaries with the Neumann boundary condition on $f_a$, 
the smallest eigenvalue $\lambda_1$ is zero and $f_1 (x)$ is a constant function. By the normalization condition, $f_1 (x)$ is 
given by $V^{-1/2}$, where $V$ is the volume, $V=\int_{\varSigma} \sqrt{\gamma}\mathd^d x$.

\item The eigenfunctions $\tilde{f}_a$ with a damping factor is defined by $\mathe^{-\alpha\lambda_a}f_a(x)$ and $Q_{abc}$ is
\begin{align*}
Q_{abc} (\alpha)&=A\int_{\varSigma}\tilde{f}_a(x)\tilde{f}_b(x)\tilde{f}_c(x)\sqrt{\gamma} \mathd^d x.\\
&=A\mathe^{-\alpha(\lambda_a+\lambda_b+\lambda_c)}\int_{\varSigma} f_a(x)f_b(x)f_c(x) \sqrt{\gamma}\mathd^d x.
\end{align*}
The real constant $A$ can be taken freely, but in this section, we assume $A=1$.

\end{enumerate}

Since the constancy of $f_1$ is very useful in the analysis, in the following discussion we consider the case 
where $(\varSigma,\gamma)$ is either a closed manifold or has boundaries with the Neumann boundary condition on $f_a$.
Note also that $\lambda_1=0$ follows in such cases.

\subsection{Discretization of integration}
\label{subsec:int}
In this subsection, we will consider discretizing the integration on $\varSigma$ and its expression by summation. 
Let us consider a set $\{p_i\in\varSigma | i = 1, 2, \ldots, R \}$ of $R$ points which are more or less uniformly distributed on $\varSigma$.
The Voronoi cell of $p_i$ is defined by
\begin{align*}
\sigma_i&=\{p\in\varSigma | \ell(p, p_i)\leq\ell(p, p_j)\mbox{ for all }j\},
\end{align*}
where $\ell(p, q)$ is the geodesic distance between $p$ and $q$. 
By this definition, the points equidistant from $p_i$ and $p_j$ are included in both $\sigma_i$ and $\sigma_j$. Since $\bigcup_i\sigma_i=\varSigma$ holds, the full volume $V$ of $\varSigma$ is the sum of the volume $V_i$ of $\sigma_i$. 
With these preparations above, we consider the integration of a scalar field $\phi(x)$ over $\varSigma$. We assume that $\phi (x)$ 
does not fluctuate violently and the Voronoi cells are sufficiently small (i.e., $R$ is large). 
Then the integration can be asymptotically approximated by
\begin{align*}
\int_{\varSigma}\phi(x)\sqrt{\gamma}\mathd^d x&=\sum_{i = 1}^R\int_{\sigma_i}\phi (x) \sqrt{\gamma}\mathd^d x\\
&\simeq\sum_{i=1}^R \phi(p_i)\int_{\sigma_i}\sqrt{\gamma}\mathd^dx\\
&=\sum_{i=1}^R\phi(p_i)V_i,
\end{align*}
where the average value of $\phi (x)$ on $\sigma_i$ has been approximated by $\phi (p_i)$, the value on the representative point $p_i$. 
This approximation transforms the integration and the summation into each other.

\subsection{Tensor rank decomposition and Laplacian}
As discussed in Section~\ref{subsec:intq}, the tensor $Q$ is defined by the integration over a Cauchy hypersurface $\varSigma$. 
By using the discussion in \ref{subsec:int}, $Q$ can be approximated by a sum,
\begin{align*}
Q_{abc}&=\int_{\varSigma} \tilde{f}_a(x)\tilde{f}_b(x)\tilde{f}_c(x)\sqrt{\gamma}\mathd^d x\\
&\simeq \sum_{i = 1}^R \tilde{f}_a(p_i)\tilde{f}_b(p_i)\tilde{f}_c(p_i)V_i.
\end{align*}
Then, by defining $\phi_a^i=\tilde{f}_a(p_i)V_i^{1/3}$, we get an expression which has the form of the tensor rank decomposition:
\begin{align}
Q_{abc}&\simeq\sum_{i=1}^R \phi_a^i\phi_b^i\phi_c^i. \label{rank decomp.}
\end{align}
This means that the discretization of the integral representation of $Q$ corresponds to the tensor rank decomposition of $Q$. 

Finally, we will explain the method to get the geometric information from $\phi_a^i$. 
Let us define $K^{ij}=\phi_a^i\phi_a^j$. 
The following approximation shows that this quantity is related to the ``fuzzy'' heat kernel $Z_N (p, q;\alpha)$:
\s[
K^{ij}&=\tilde{f}_a(p_i)\tilde{f}_a(p_j)V_i^{1/3}V_j^{1/3}\\
&=\sum_{a=1}^N \mathe^{-2\alpha\lambda_a}f_a (p_i)f_a(p_j)V_i^{1/3}V_j^{1/3}\\
&\simeq Z_N(p_i,p_j;2\alpha)V_i^{1/3}V_j^{1/3},
\label{eq:kzrel}
\s]
where $Z_N(p,q ;\alpha)$ is defined by
\begin{align}
Z_N(p,q;\alpha)&=\sum_{a=1}^N \mathe^{-\alpha\lambda_a} f_a(p)f_a(q).
\label{eq:defofzn}
\end{align}
The limit $N\rightarrow\infty$ of $Z_N(p,q;\alpha)$ agrees with the standard heat kernel.
We note that $Z_N$ satisfies the heat equation
\begin{align*}
\frac{\partial}{\partial\alpha} Z_N (x,y;\alpha)&=\Delta_x Z_N(x,y;\alpha) .
\end{align*}
The result \eqref{eq:kzrel} implies that, if we know $V_i$, the eigenvalues $\mathe^{-2\alpha\lambda_a}$ of $Z_N(p_i,p_j;2\alpha)$
can be computed from $K^{ij}$. 
So the next question is whether $V_i$ can be calculated. To answer it, let us first derive the following equation,
\begin{align*}
\sum_{j=1}^R K^{ij}V_j^{2/3}&=\sum_{a=1}^N \mathe^{-2\alpha\lambda_a} f_a(p_i)V_i^{1/3}\sum_{i=1}^R f_a(p_j)V_j\\
&\simeq \sum_{a=1}^N \mathe^{-2\alpha\lambda_a} f_a(p_i)V_i^{1/3}\int_{\varSigma} f_a(x)\sqrt{\gamma}\mathd^d x\\
&=\sum_{a=1}^N \mathe^{-2\alpha\lambda_a} f_a(p_i)V_i^{1/3}\delta_{a,1}V^{1/2}\\
&=V_i^{1/3}.
\end{align*}
This means that $V_i$ is one of the solutions of the equation,
\begin{align}
\sum_{j=1}^R K^{ij}X_j^{2/3}&=X_i^{1/3},
\label{volume formula}
\end{align}
with $^\forall X_i \geq 0$.  At this stage, since \eq{volume formula} has been derived with approximations, 
it is not clear whether \eq{volume formula} has always a solution for a given $\phi_a^i$, or, if so, whether a solution can be uniquely
identified with $V_i$. These questions will be discussed in the following subsection.

\subsection{Tensor eigensystems}

The tensor eigenvector $(v_a) \in\mathbb{C}^N\backslash\{ 0 \}$ and the associated eigenvalue $\lambda\in\mathbb{C}$ of 
a symmetric tensor $Q_{abc}$ are defined by \cite{qi1,qi2,tensoreigen}
\begin{align}
Q_{abc}v_bv_c&=\lambda v_a.\label{eigensystem}
\end{align}
If $(v, \lambda)$ is an eigensystem, its scaled pair $(tv, t\lambda)$ with any $t\in\mathbb{C}\backslash\{ 0 \}$ is also an eigensystem. 
This property introduces the natural identification $(v, \lambda) \sim (tv, t\lambda)$. 
The total number of the complex eigenvectors of \eqref{eigensystem} under this equivalence class 
has been shown to be $2^N-1$, if it is finite \cite{tensoreigen}.

We are interested in real solutions for real $Q$. Let us first point out that there always exists at least one real solution.
The reason is as follows. Unless $Q=0$, $Q_{abc}v_av_bv_c$ has a maximum positive value on the sphere $|v|=1$.
By applying the method of Lagrange multiplier, the equation for the maximum leads to the same equation as  \eq{eigensystem},
where $\lambda>0$ for the positive maximal value. 

Now, to discuss the relation between \eq{eigensystem} and \eq{volume formula}, let us consider a solution of \eq{eigensystem}
with $\lambda\neq 0$. One can always take $(v, \lambda) \sim (v', 1)$ by taking $t=1/\lambda$, and put \eq{eigensystem} 
into the form, 
\begin{align}
Q_{abc}v_b v_c=v_a.\label{eigensystem2}
\end{align}
By putting the tensor rank decomposed form of the tensor $Q$ \eqref{rank decomp.} into \eq{eigensystem2}, we obtain
\begin{align}
\sum_{j=1}^R \phi_a^j(v^j)^2=v_a,
\label{contract}
\end{align}
where $v^j=\phi_a^j v_a$.
By multiplying the both sides of \eqref{contract} by $\phi_a^i$ and taking the contraction, we get
\begin{align}
\sum_{j=1}^R K^{ij}(v^j)^2=v^i.\label{volume formula2}
\end{align}
If ${}^\forall v^i\geq 0$, one can obtain the solution to \eqref{volume formula} by identifying $X_i=(v^i)^3$. 
In other words, there are as many solutions to \eqref{volume formula} as the eigenvectors of $Q$ satisfying $^\forall v^i\geq 0$.

It does not seem to be an easy problem to find the general conditions under which 
we can find the eigenvectors satisfying $^\forall v^i\geq 0$ and can uniquely identify $X_i$.  
Therefore, below, we will restrict ourselves to just explicitly showing that, when $R$ is large enough and $\alpha$ is small enough, 
the eigensystem equation \eq{eigensystem2} has a solution which gives $X_i\simeq V_i$.

From $\tilde f_1(x)=V^{-1/2}$ and $\lambda_1=0$, we get a formula for $Q_{1ab}$,
\s[
Q_{1ab}&=\mathe^{-\alpha(\lambda_a+\lambda_b)}V^{-1/2}\int_\varSigma f_a(x)f_b(x)\sqrt{\gamma}\mathd^d x\\
&=V^{-1/2}\mathe^{-2\alpha\lambda_a}\delta_{ab}.
\label{eq:q1}
\s]
By using this, we can find
\begin{align*}
Q_{1ab}v_av_b&=V^{-1/2}\sum_{a\ge1}\mathe^{-2\alpha\lambda_a}v_a^2\\
&=V^{-1/2}v_1^2+V^{-1/2}\sum_{a\ge2}\mathe^{-2\alpha\lambda_a}v_a^2.
\end{align*}
Therefore, by using \eq{eigensystem2}, the $a=1$ component of \eqref{eigensystem2} can be rewritten as
\begin{align}
\sum_{a\ge2}\mathe^{-2\alpha\lambda_a}v_a^2=v_1(V^{1/2}-v_1).\label{a=1}
\end{align}
For any real vector $v_a$, the LHS of \eqref{a=1} cannot be negative. 
This implies $0\le v_1\le V^{1/2}$, and in particular, if $v_1=0$ or $v_1=V^{1/2}$, we have $v_a=0$ for all $a\ge2$.
In fact, $(v_a)=(V^{1/2},0,0,\ldots)$ is a solution to \eq{eigensystem2}, because
\[
Q_{abc} v_b v_c=Q_{a11}v_1v_1=0,\ \hbox{for }a\geq 2
\]
from \eq{eq:q1}.
We note that the inner product of $\tilde f_a(x)$ and the non-zero solution $(v_a)=(V^{1/2},0,0,\ldots)$ is equal to 1. 
Correspondingly, $v^i=\phi^i_a v_a\simeq V_i^{1/3}\tilde f_a(p_i)v_a=V_i^{1/3}$, meaning $X_i\simeq V_i$. 
This implies that, if $R$ is sufficiently large, there will always be such a solution to the equation \eqref{volume formula2}.

Furthermore, if we also assume $\alpha\gg 1/\lambda_2$, any eigenvector of $Q$ gives $X_i\simeq V_i$.
This can be shown in the following way.
First of all, we know 
\begin{align*}
Q_{abc}v_bv_c&=2Q_{ab1}v_bv_1+\sum_{b,c\ge2}Q_{abc}v_bv_c\\
&=2V^{-1/2}\mathe^{-2\alpha\lambda_a}v_av_1+\sum_{b,c\ge2}Q_{abc}v_bv_c
\end{align*}
for $a\ge2$.
So the $a\ge 2$ components of \eqref{eigensystem2} can be rewritten as
\begin{align}
\sum_{b,c\ge2}Q_{abc}v_bv_c=v_a(1-2V^{-1/2}\mathe^{-2\alpha\lambda_a}v_1).\label{a>1}
\end{align}
Since the RHS of \eqref{a=1} is less than or equal to $V/4$, we can see that $|v_a|\le \mathe^{\alpha\lambda_a}V^{1/2}/2$ for all $a\ge2$.
Based on this, the absolute value of LHS of \eqref{a>1} can be bounded from above in the following way,
\begin{align*}
\left|\sum_{b,c\ge2}Q_{abc}v_bv_c\right|&=\left|\sum_{b,c\ge2}Q_{abc}^{(\alpha=0)}\mathe^{-\alpha(\lambda_a+\lambda_b+\lambda_c)}v_bv_c\right|\\
&\le\mathe^{-\alpha\lambda_a}\sum_{b,c\ge2}\left|Q_{abc}^{(\alpha=0)}\right|\left|\mathe^{-\alpha\lambda_b}v_b\right|\left|\mathe^{-\alpha\lambda_c}v_c\right|\\
&\le\mathe^{-\alpha\lambda_a}\sum_{b,c\ge2}\frac{V}{4}\left|Q_{abc}^{(\alpha=0)}\right|.
\end{align*}
Clearly, this asymptotically approaches zero for $\alpha\gg 1/\lambda_2$.
On the othe hand, the absolute value of RHS of \eqref{a>1} is asymptotically equal to $|v_a|$.
Combining the above results, we find that $v_a$ approaches zero for all $a\ge2$, and by \eqref{a=1}, $v_1$ must be equal to $V^{1/2}$.
This means that the solution approaches the above solution $(v_a)=(V^{1/2},0,0,\ldots)$ or it is the above solution itself.
By the same argument as above, the solution gives $v^i\simeq V_i^{1/3}$, meaning $X_i\simeq V_i$.

\section{Geometric properties of the outer bunch}
\label{sec:geometry}
In this section, we will show that, in the classical phase, the pattern of the set of order parameters \eq{eq:phi2outer} forms an object
which has the geometry expected from the Lie group invariance of $Q$. 
We will demonstrate this for the $Q$'s constructed in Section~\ref{sec:symq}.

\begin{figure}
\begin{center}
\includegraphics[width=7cm]{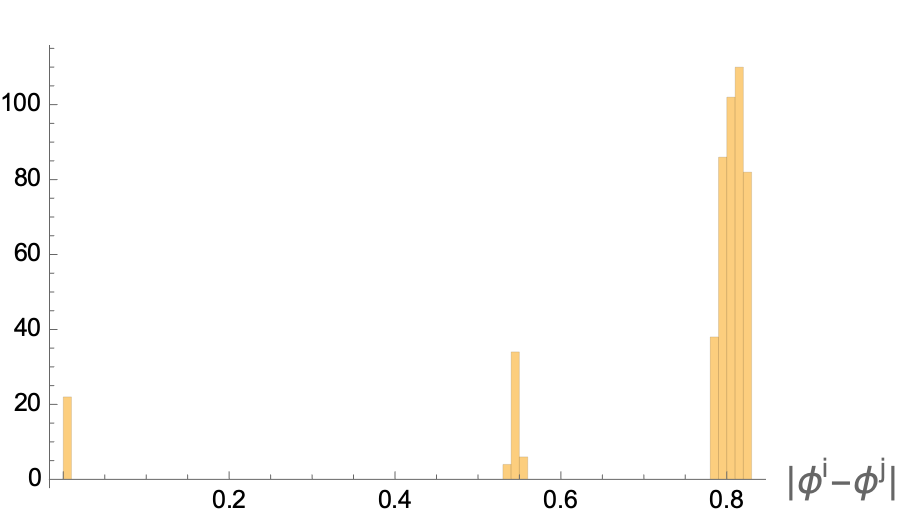}
\hfil
\includegraphics[width=4cm]{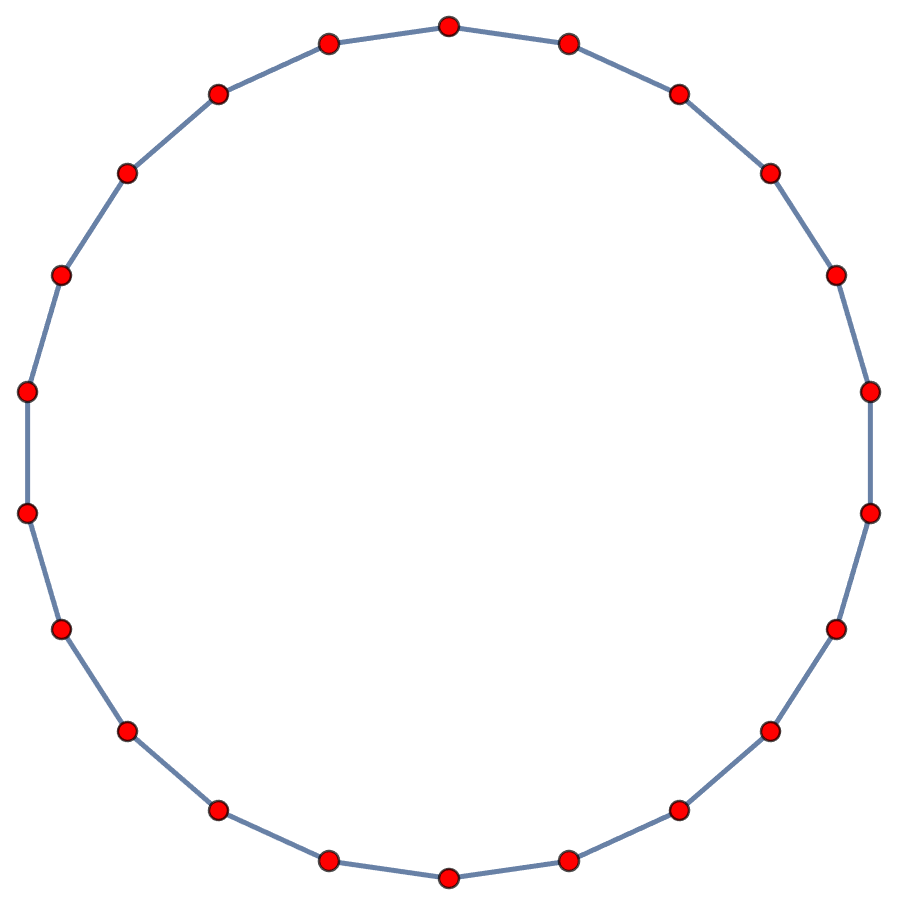}
\caption{Left: The histogram of the distances between $\phi^i$ and $\phi^j$ in the outer bunch. 
The leftmost peak represents the zero distance 
between $\phi^j$ themselves, and the peak around 0.55 represents the distances between the nearest neighbors. 
The parameters are taken $N=15\ (\Lambda=7)$, 
$\lambda=10^7$, and $\alpha=0.5$. 
Right: Linking the nearest neighbors makes an $S^1$.
}
\label{fig:s1}
\end{center}
\end{figure}

\begin{figure}
\begin{center}
\includegraphics[width=7cm]{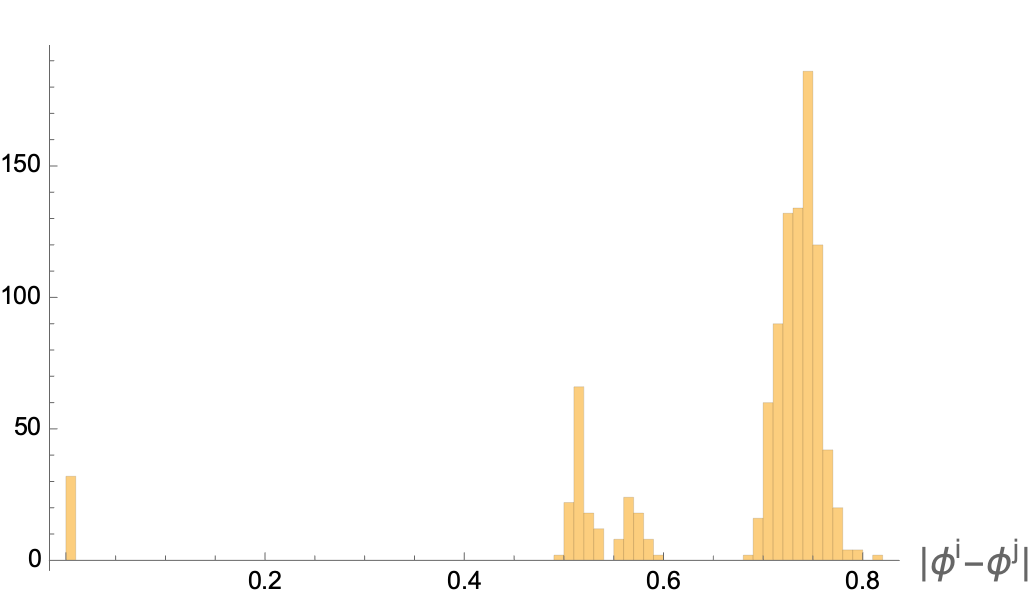}
\hfil
\includegraphics[width=4cm]{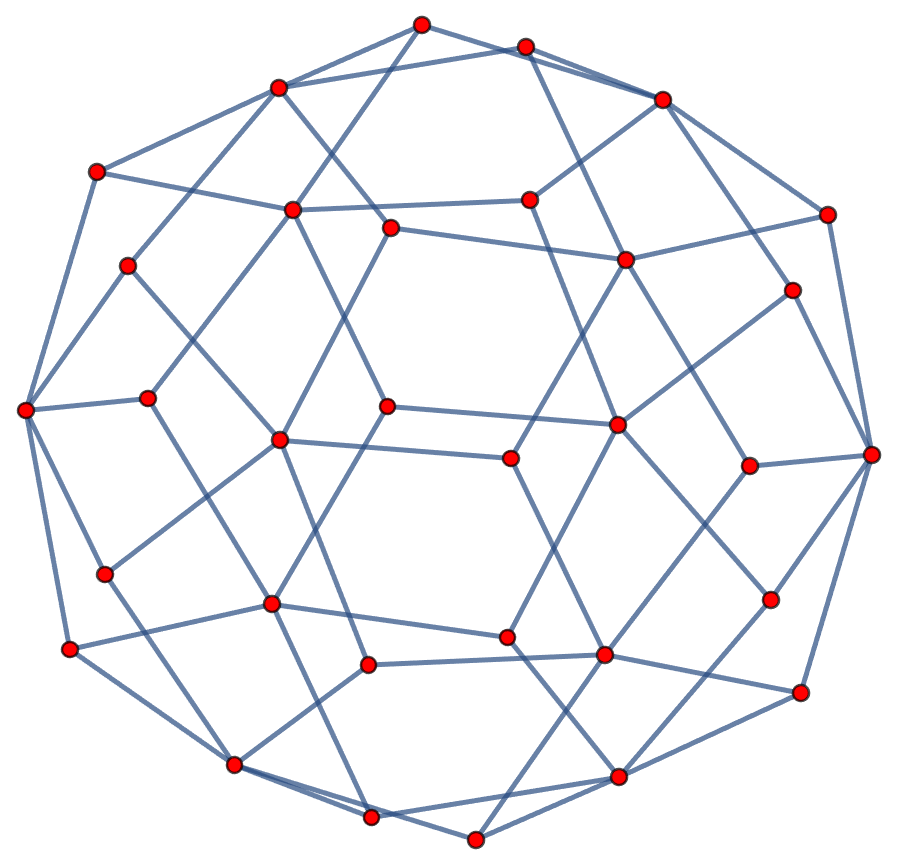}
\caption{Left: The histogram made in the same way as in Figure~\ref{fig:s1} for $Q^{SO(3)}$ with $N=16\ (\Lambda=3)$, 
$\lambda=10^7$, and $\alpha=0.5$. 
Right: Linking the nearest neighbors makes an $S^2$.
}
\label{fig:s2}
\end{center}
\end{figure}

\subsection{Topology}
\label{sec:topology}
Let us first consider $Q^{SO(2)}$. As shown in Section~\ref{sec:twophases}, the inner products
$\phi^i_a \phi_a^j$ within the outer bunch almost take constant values in the classical phase.
Therefore it does not lose generality to pick up one arbitrary sample of 
$\phi$ from a MC sequence as a representative. 
Then we collect the $\phi^i$'s in the outer bunch according to their sizes. 
For the case in Figure~\ref{fig:s1}, the number of $\phi^i$'s in the outer bunch is 22 out of the total $R=76$ (Namely,
\eq{eq:rval} for $N=15$.). 
Then, as shown in the left panel of Figure~\ref{fig:s1}, we compute all the mutual distances between $\phi^i$'s in the outer bunch, 
identify the nearest neighbor pairs of $\phi^i$'s, and connect them with lines. After the process we find an $S^1$ as in the right panel. 

This can be done for $Q^{SO(3)}$ as well, and we find $S^2$ as in Figure~\ref{fig:s2}. 
The same thing can also be done for $Q^{SO(4)}$, but the corresponding figure is not shown, since 
it is difficult to see the topology of $S^3$ on a two-dimensional sheet.

\subsection{Spectra of Laplacian -- geometry}
An efficient way of detecting the geometry of an emergent space is to define a Laplacian on it and study the spectra.
We will give two definitions and apply them to the configurations of the outer bunches obtained for $Q^{SO(n+1)}$.
In both cases, we obtain spectra consistent with the Laplacian on $S^n$, supporting 
the topological study in Section~\ref{sec:topology}.

The first definition comes from Section~\ref{sec:trd}. Since the construction of $Q^{SO(n+1)}$ is based on the way described there, 
it would be reasonable to employ the spectra of Laplacian derived from the discussions there.  
The quantity defined in \eq{eq:defofzn} can be regarded as an operator with eigenvalues $\mathe^{-2 \alpha \lambda_a}$,
where $\lambda_a$ are the spectra, in the following sense:
\s[
\int_{\varSigma}  \mathd^d q\,  \sqrt{\gamma}\, Z_N(p,q;2 \alpha) f_b(q)&=\int_{\varSigma} \mathd^d q \sqrt{\gamma}\, \sum_{a=1}^N
\mathe^{-2 \alpha \lambda_a} f_a(p) f_a(q)  f_b(q) \\
&= \mathe^{-2\alpha \lambda_b} f_b(p),
\s]
where we have used the orthonormal condition \eq{eq:normf}.
After discretization, this is expressed by a matrix,
\[
Z_N(p_i,p_j;2 \alpha) V_j \simeq  V_i^{-1/3} K^{ij} V_j^{2/3},
\label{eq:kvnaive}
\]
by using \eq{eq:kzrel}. The expression on the righthand side is inconvenient, because
the matrix is not symmetric. By performing a similarity transformation by $V_i^{1/2}$ and $V_j^{-1/2}$ 
to \eq{eq:kvnaive} respectively from the left and the right, we obtain an equivalent eigenvalue problem with a symmetric matrix,
\[
V_i^{1/6} K^{ij} V_j^{1/6}.
\label{eq:mat1}
\]

For convenience, let us relabel the $\phi^i$ in the outer bunch as $\phi^i\  (i=1,2,\ldots,\tilde R)$ without loss of generality,
where $\tilde R$ is the total number of $\phi^j$ in the outer bunch.
When $\tilde R>N$, which turns out to be our case, the matrix \eq{eq:mat1} has at least $\tilde R-N$ zero eigenvalues, because 
\eq{eq:mat1} has the product form of two matrices, $(V_i^{1/6}  \phi^i_a)( \phi^j_a V_j^{1/6})$, 
where the vector space associated to the lower index
in the middle has the smallest dimension $N$. However, we are only interested in the positive eigenvalues $e^{-2\alpha \lambda_a}$, and 
a convenient way to extract these positive eigenvalues is to consider the other way of the composition of the two matrices,
\[
L_{ab}:=\sum_{i=1}^{\tilde R} \phi^i_a V_i^{1/6} V_i^{1/6}  \phi^i_b= \sum_{i=1}^{\tilde R} \phi_a^i V_i^{1/3} \phi_b^i.
\label{eq:L}
\] 
The eigenvalues of the matrix $L$ give the spectra $e^{-2 \alpha \lambda_a}$. Here the values of $V_i$ are determined by
solving \eq{volume formula}.

\begin{figure}
\begin{center}
\includegraphics[width=5cm]{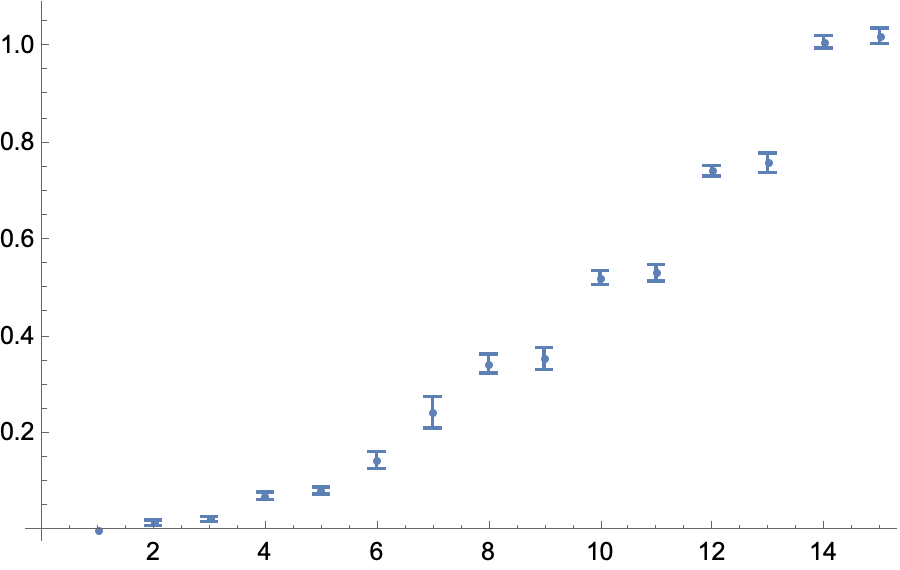}
\hfil
\includegraphics[width=5cm]{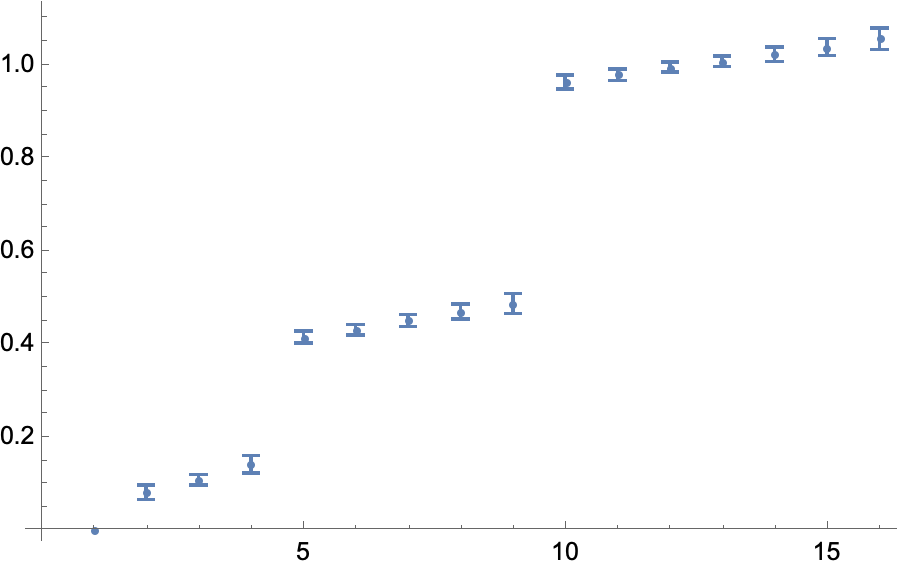}
\hfil
\includegraphics[width=5cm]{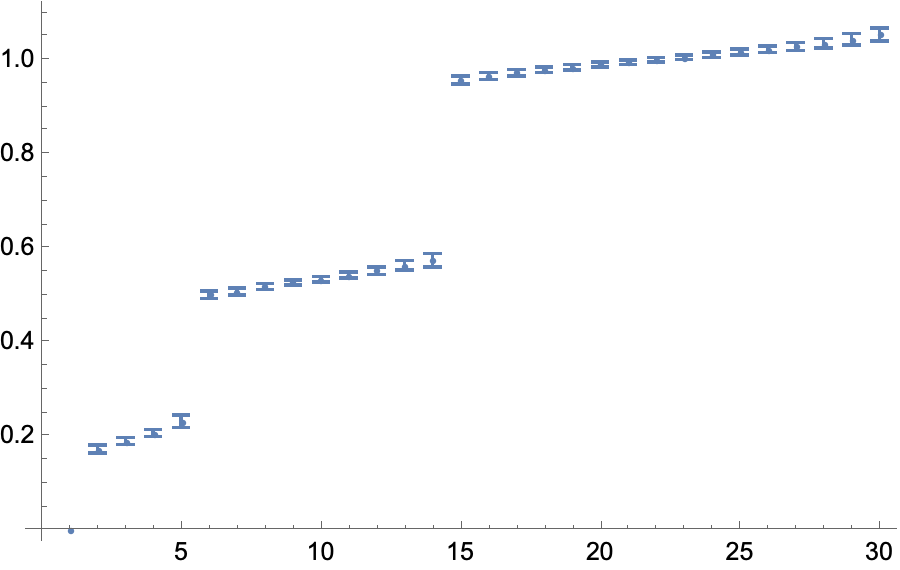}
\caption{The minus of the logarithm of the eigenvalues of $L$ in \eq{eq:L} are plotted in order along the horizontal axis.
The values are shifted vertically so that the smallest values be zero.   
$Q^{SO(n+1)}\ (n=1,2,3)$ from the left to the right panels. 
The parameters are the same as in Figures~\ref{fig:so2tran} to \ref{fig:so4tran} with $\lambda=10^{-7}$.
$\tilde R=22,\ 32,\ 82$ respectively from the left to the right cases. 
500 samples from the sequences are statistically analyzed for each case. 
}
\label{fig:speclap}
\end{center}
\end{figure}

In Figure~\ref{fig:speclap}, the minus of the logarithm of the eigenvalues of $L$ are plotted. The degeneracies and the values
are consistent with what are expected for $S^n\ (n=1,2,3)$. A curious matter is that the 5th and 6th eigenvalues 
in the $S^1$ case (the left panel) look deviated compared to the others. 
We have performed other runs, but this seems universal for all the data of this case. 
This may imply an instability of $S^1$, but we do not currently have a definite explanation.

\begin{figure}
\begin{center}
\includegraphics[width=5cm]{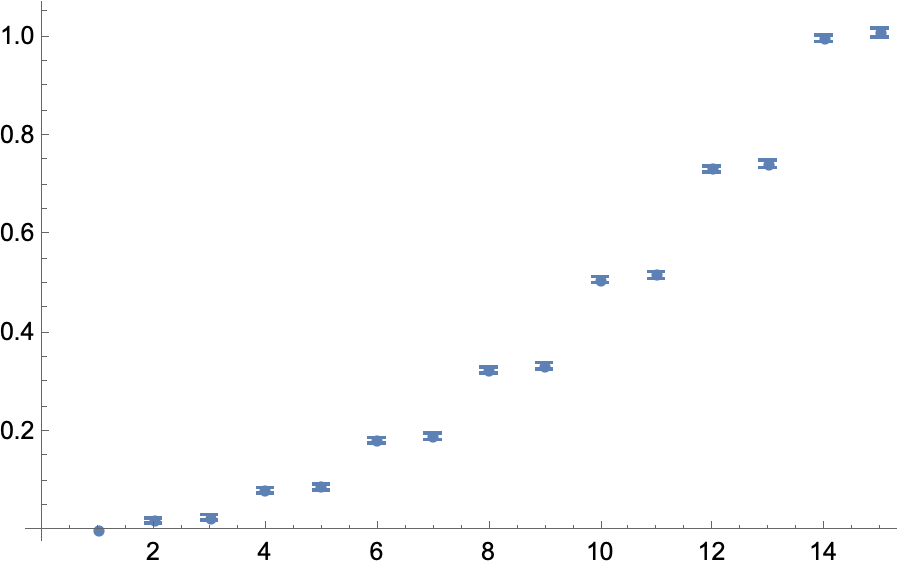}
\hfil
\includegraphics[width=5cm]{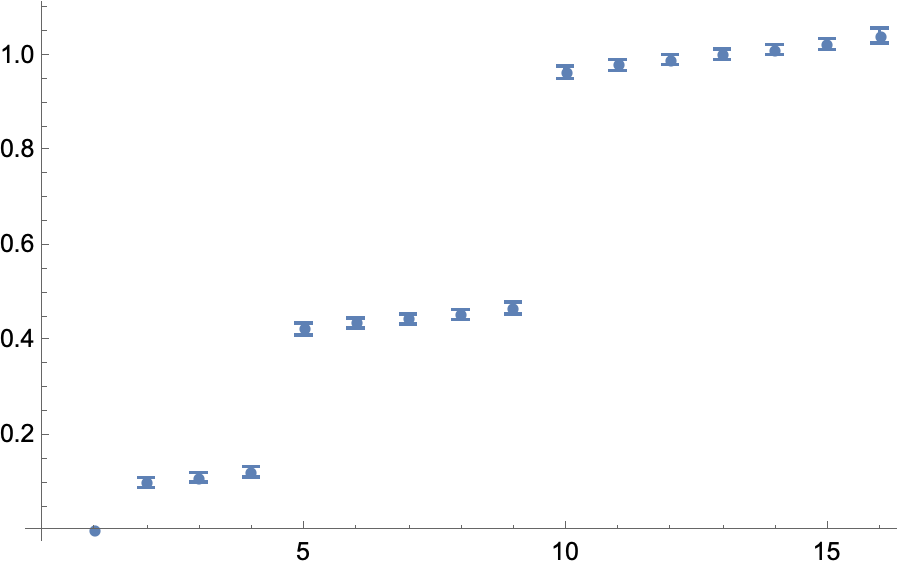}
\hfil
\includegraphics[width=5cm]{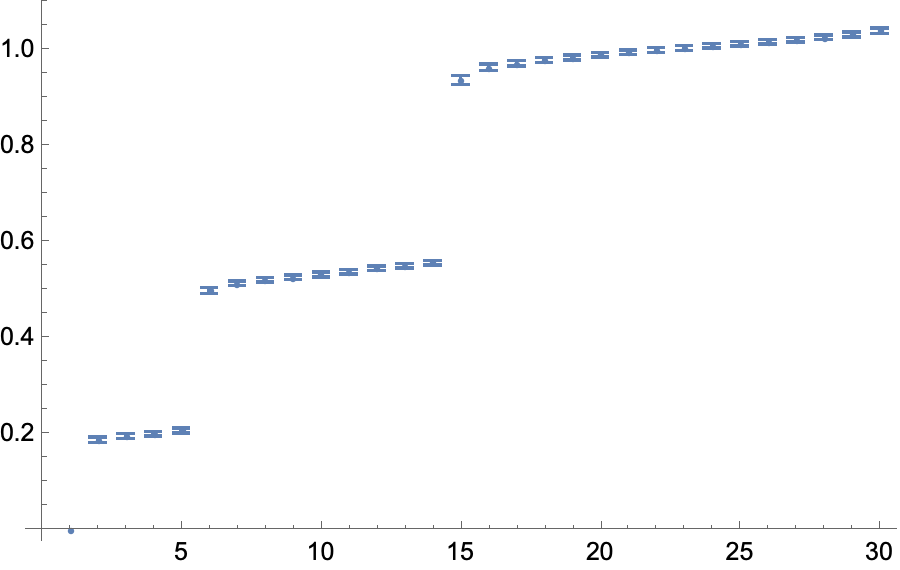}
\caption{The same for $\tilde L$ in \eq{eq:tildeL} as in Figure~\ref{fig:speclap}.}
\label{fig:speclapd}
\end{center}
\end{figure}

Another definition of the Laplacian is to skip the procedure to determine $V_i$:
\[
\tilde L_{ab}=\sum_{i=1}^{\tilde R} \phi_a^i\phi_b^i.
\label{eq:tildeL}
\]
This definition more clearly incorporates the symmetry of the system, namely, the replica symmetry.
Since this definition does not require positive solutions for $V_i$, 
it can be used for any circumstances, even in the quantum phase, where there would be quite small chances of getting 
positive solutions for $V_i$. On the other hand, 
this definition relies on the assumption that the $\phi^i$'s take reasonable values from the dynamics, 
while the former definition can absorb the ambiguities related to the discretization discussed in Section~\ref{sec:trd}.
Currently, we are not sure of which one is better and must wait for future study, 
but we would expect that these two definitions would produce more or less identical results in the large $N$ limit in the classical phase.
Figure~\ref{fig:speclapd} shows the spectra obtained from $\tilde L$, supporting the expectation.

\section{Deformations of $Q$}
\label{sec:deform}
In this section, we study two kinds of deformations of $Q$ from those given in Section~\ref{sec:symq}.

\subsection{Dependence on representations}
In the construction of Lie group invariant $Q$'s in Section~\ref{sec:symq}, the vector spaces associated to 
the indices of $Q$ are given by direct sums of some irreducible representation spaces of the Lie groups.
The sets of irreducible representations considered there are taken successively from the trivial representation to the one indexed 
by a cutoff $\Lambda$. 
This is a natural choice from the physical point of view, since $\Lambda$ can naturally be
related to a short-distance cutoff in a space. In this section, we consider some unnatural choices by 
dropping a few of the intermediate representations.

\begin{figure}
\begin{center}
\includegraphics[width=7cm]{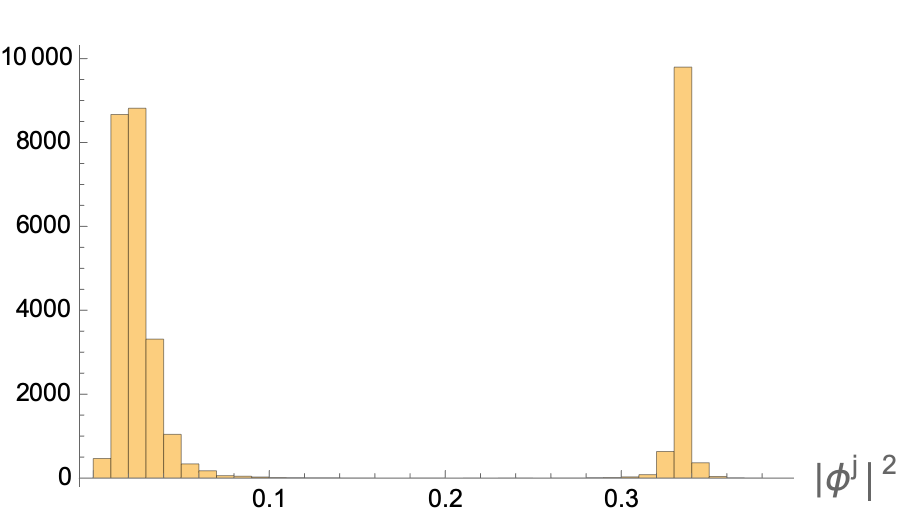}
\hfil
\includegraphics[width=7cm]{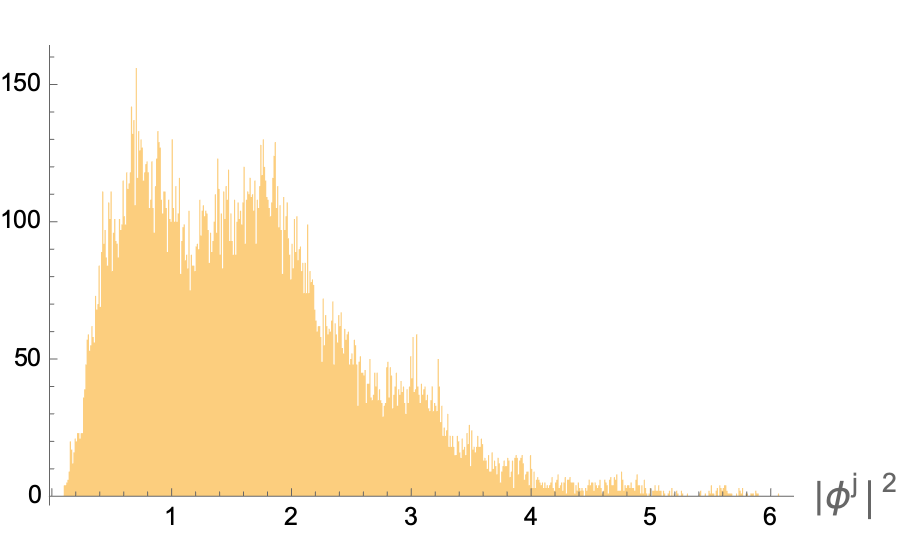}
\caption{Dependence of phases on representations. The examples are for $Q^{SO(2)}$ with $\Lambda=7$,
$\lambda=10^7$, and $\alpha=0.5$. 
Left: The Histogram of $|\phi^j|^2$ when $p=0$ is dropped. Right: $p=0,1$ are dropped.}
\label{fig:omitso2}
\end{center}
\end{figure}

\begin{figure}
\begin{center}
\includegraphics[width=7cm]{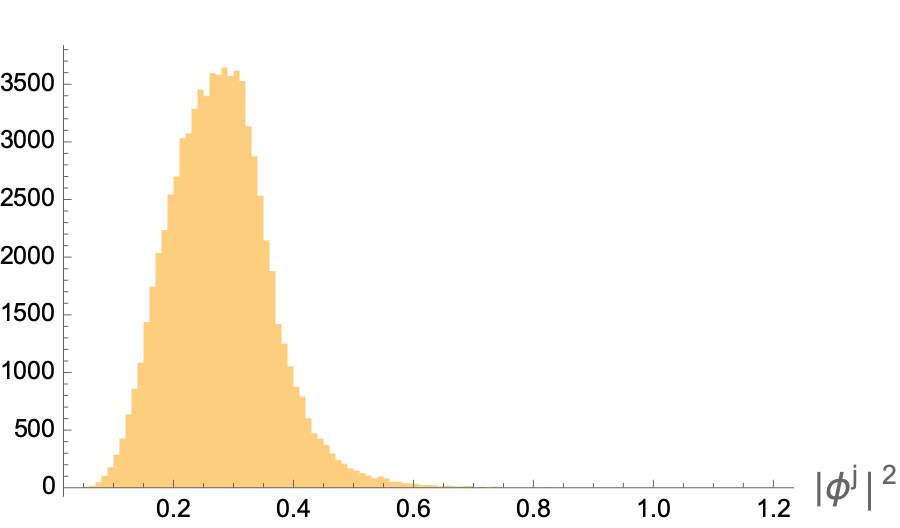}
\hfil
\includegraphics[width=7cm]{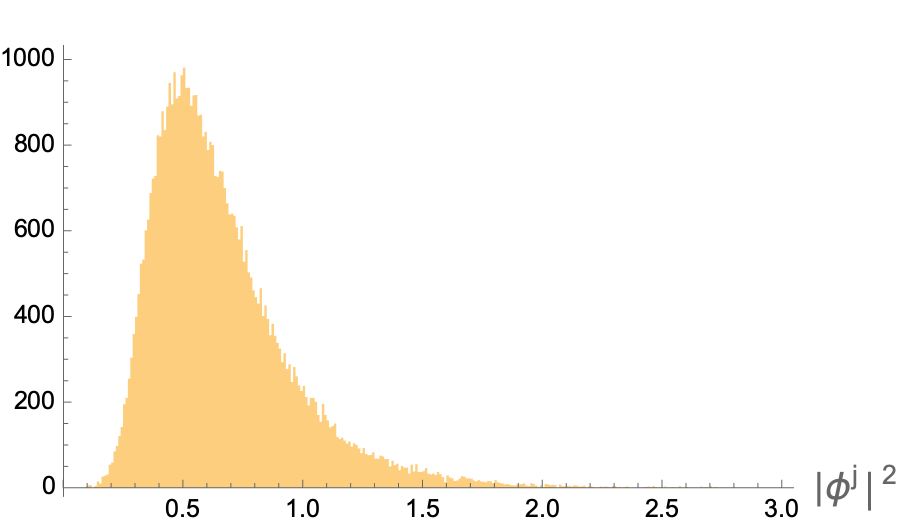}
\caption{Dependence of phases on representations. The examples are for $Q^{SO(3)}$ with $\Lambda=3$, $\lambda=10^7$, 
and $\alpha=0.5$. 
Left: $l=0$ is dropped. Right: $l=0,1$ are dropped.}
\label{fig:omitso3}
\end{center}
\end{figure}

In Figure~\ref{fig:omitso2}, we consider $Q^{SO(2)}$ with $\Lambda=7$, $\lambda=10^7$, and $\alpha=0.5$. 
As is shown in Figure~\ref{fig:classreg},
this case is in the classical phase, if we consider the full representations, $p=0,1,\ldots,\Lambda=7$, namely $N=15$. 
The left panel of Figure~\ref{fig:omitso2} shows the histogram of $|\phi^j|^2$ when
we drop the $p=0$ representation. This is still in the classical phase. However, when we drop $p=0,1$,
the phase becomes the quantum as shown in the right panel.
In Figure~\ref{fig:omitso3}, we consider $Q^{SO(3)}$. Even dropping $l=0$ 
changes the phase to the quantum.

One would suspect that the change to the quantum phase is merely caused by the reduction of $N$ due to dropping. However,
dropping $p=0,1$ for $Q^{SO(2)}$ above corresponds to $N=12$. According to Figure~\ref{fig:classreg}, the case with 
$N=12$ and $\lambda=10^7$ is in between the quantum and the classical phases, but the histogram on the right panel of
Figure~\ref{fig:omitso2} shows the clear characteristic of the quantum phase.
Therefore we see that dropping intermediate representations
makes the classical phase less likely.
It is interesting that the most physically natural set of representations, namely, successively taken from 
the trivial representation to a cut-off, makes the classical phase most likely. 

\subsection{Breaking Lie group invariance of $Q$} 
\label{sec:offsym}
In this subsection, we break Lie-group invariance of $Q$ by
\[
Q=(Q^{SO(n)}+z\, Q^B)/\sqrt{1+z^2},
\] 
where $z$ is a deformation parameter, $Q^{SO(n)}$ is given in Section~\ref{sec:symq}, and 
$Q^B$ is a tensor which breaks the Lie group invariance. 

There are too many possible such $Q^B$'s, and it is not feasible to study all of them. 
We therefore consider merely one such $Q^B$, which seems meaningless enough to consider it as reflecting the general 
effect. One way to give $Q^B$ is to assign random values, but this would have the problem of reproducing
the result. So, we rather consider a meaningless function to generate $Q^B$. More precisely, we consider   
\[
Q^B_{abc}=const. \cdot \left\{ 
\begin{array}{cl}
\cos\left[0.1(a+b+c)\right]  & \hbox{if } Q^{SO(n)}_{abc}=0 \\
0 & \hbox{otherwise}
\end{array}
 \right.
 ,
\]
with the normalization $const.$ for $|Q^B|=1$. This also satisfies the transversality,  $Q^{SO(n)}_{abc} Q^B_{abc}=0$.

\begin{figure}
\begin{center}
\includegraphics[width=5cm]{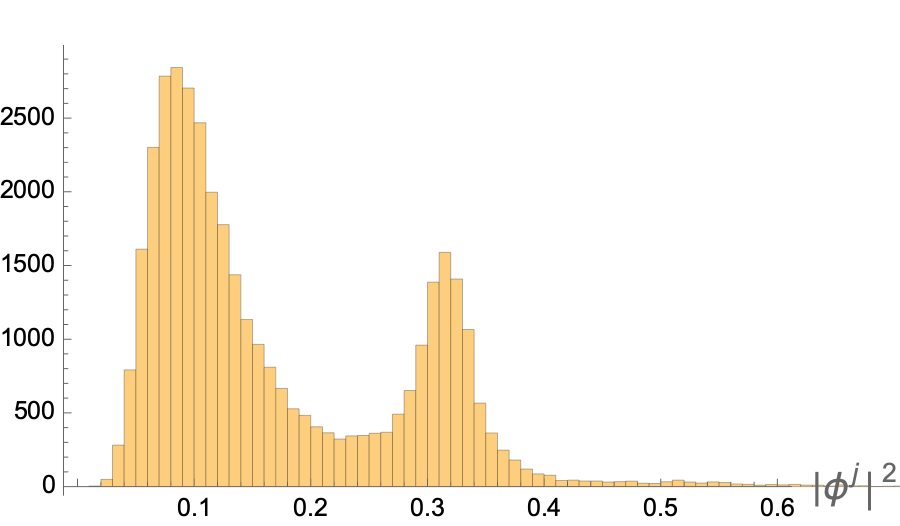}
\hfil
\includegraphics[width=5cm]{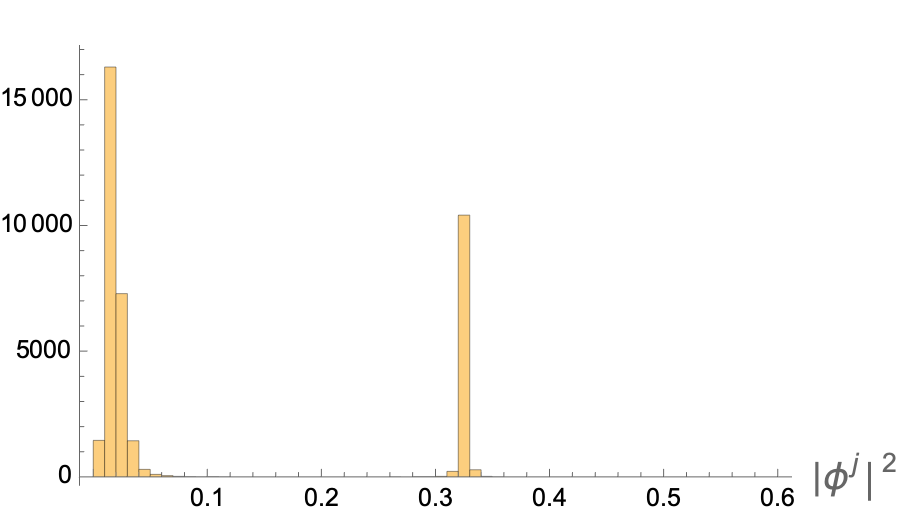}
\hfil
\includegraphics[width=5cm]{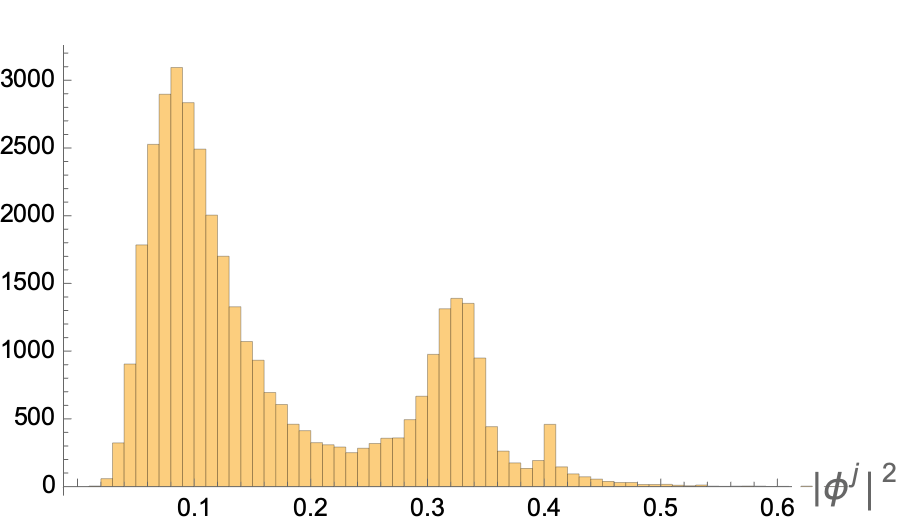}
\caption{The histograms of $|\phi^j|^2$ for $z=-24/85,0,24/85$ from the left panel to the right, respectively.
The setting is $Q^{SO(2)}$ with $N=15 \, (\Lambda=7)$, $\lambda=10^7$, and $\alpha=0.5$.
}
\label{fig:offso2}
\end{center}
\end{figure}

The result is shown in Figure~\ref{fig:offso2} for $Q^{SO(2)}$. Non-zero $z$ makes the classical phase less likely. 

We have also observed a similar effect for $Q^{SO(3)}$.
We conclude that breaking Lie group invariance makes the classical phase less likely. 

\section{Behavior of the oscillatory part} 
\label{sec:oscillatory}
In this section, we discuss the oscillatory part, 
\[
\left\langle
\prod_{i=1}^R {\rm Airy}\left(-|Q|^{2/3} (\phi^i)^2\right) \right\rangle_{\tilde Q,\lambda},
\label{eq:fixedlam}
\]
in \eq{eq:repsiq}, where $\tilde Q=Q/|Q|$. Here note that we have dropped the dependence of the coupling parameter 
on $|Q|^2$ compared to \eq{eq:repsiq}. The reason for this simplification is that changing the coupling parameter requires 
repeating Monte Carlo simulations with different couplings, and it would not be feasible to obtain results for many $|Q|$'s.  
On the other hand, 
$\eq{eq:fixedlam}$ for different $|Q|$'s can be computed by using the same sampling data of the Monte Carlo simulation.
At the end of this section we discuss some speculations about the effect of including the dependence on $|Q|^2$ back.

The fluctuations of $|\phi^j|^2$ are larger in the quantum phase than in the classical phase. 
Since the Airy function \eq{eq:actairy} in \eq{eq:fixedlam} is an oscillatory function, larger fluctuations make 
the expectation values smaller, because there are more cancellations. 
Therefore \eq{eq:fixedlam} generally takes smaller values in the quantum phase than in the classical phase.
In particular, there are more cancellations as $|Q|$ becomes larger, because of  the $|Q|$ dependence in the argument
of \eq{eq:fixedlam}.

\begin{figure}
\begin{center}
\includegraphics[width=5cm]{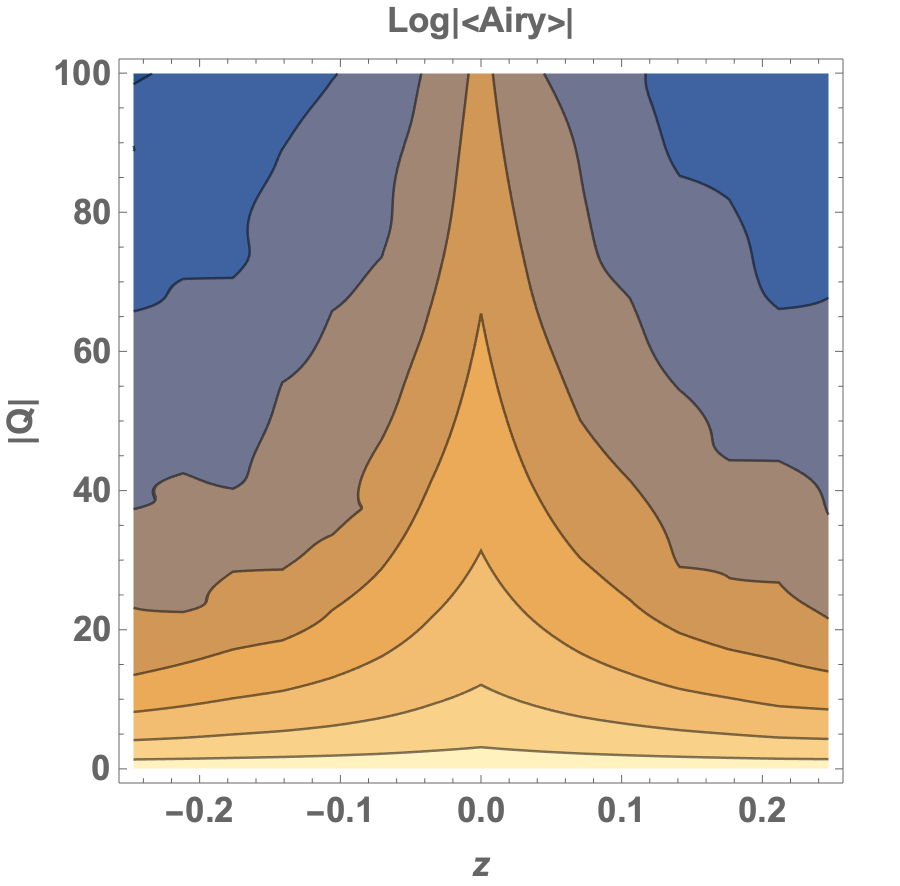}
\hfil
\includegraphics[width=5cm]{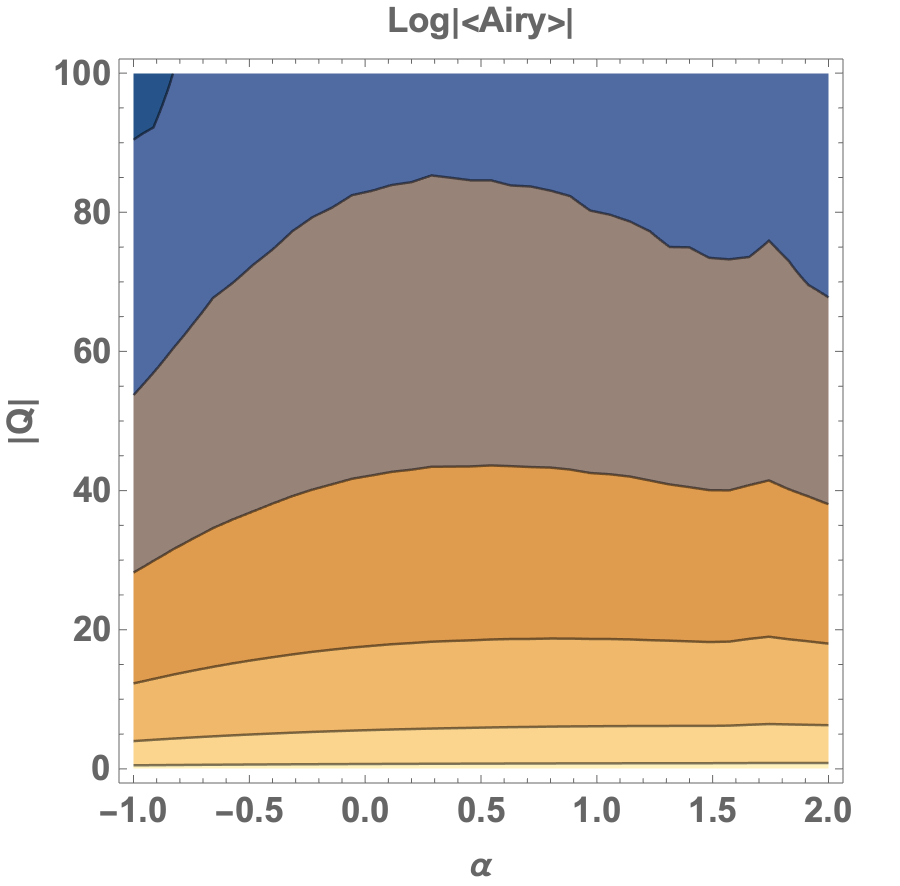}
\caption{The contour plots of $\log |\langle \prod \hbox{Airy}\rangle |$, the logarithm of the modulus of the expectation value. 
Left: A symmetry breaking case in Section~\ref{sec:offsym} with
$Q^{SO(2)}$, $N=15 \, (\Lambda=7)$, $\lambda=10^7$, and $\alpha=0.5$.
The horizontal axis represents the breaking parameter $z$, and the vertical one $|Q|$. 
Right: An invariant case with $Q^{SO(2)}(\alpha)$. $\alpha$ is varied along the horizontal axis. 
The other parameters are the same as in the left panel.}
\label{fig:logairy}
\end{center}
\end{figure}

\begin{figure}
\begin{center}
\includegraphics[width=5cm]{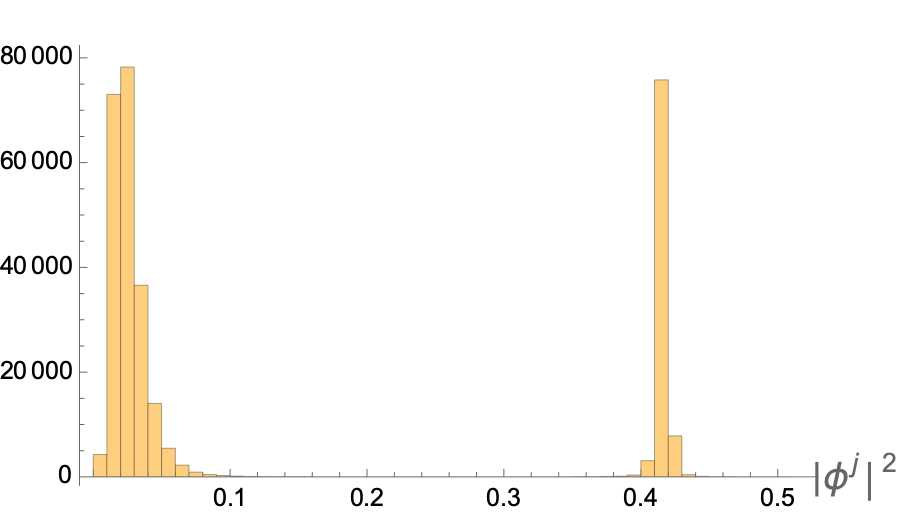}
\hfil
\includegraphics[width=5cm]{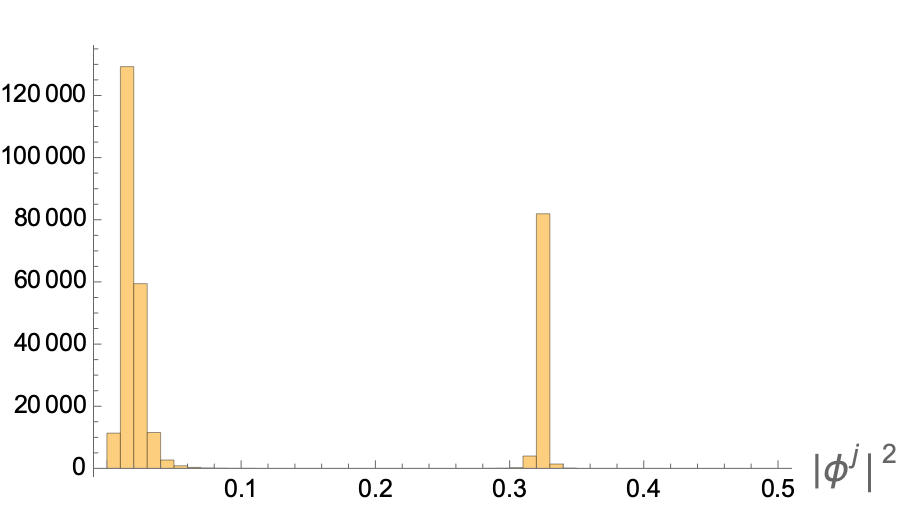}
\hfil
\includegraphics[width=5cm]{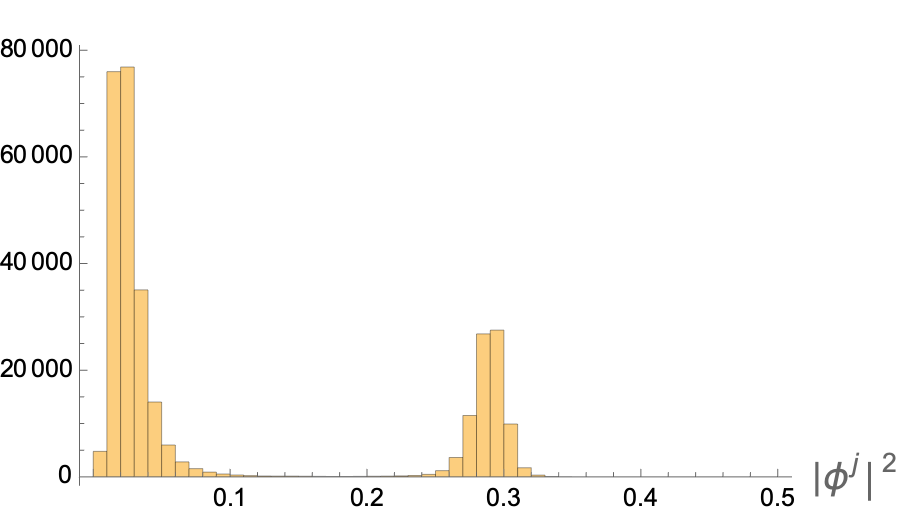}
\caption{Histograms of $|\phi^j|^2$ for $Q^{SO(2)}(\alpha)$ with $N=15 \, (\Lambda=7)$, $\lambda=10^7$, and $\alpha=$$-1$ (left), 
$19/35$ (middle), and $2$ (right).}
\label{fig:aldep}
\end{center}
\end{figure}

Figure~\ref{fig:logairy} shows the results of the expectation values from the Monte Carlo simulations. 
In the left panel, a symmetry breaking case in Section~\ref{sec:offsym} is studied. As can be seen in the figure,
the expectation value is strongly peaked around $z=0$.
This can be understood by the fact that the classical phase around $z=0$ is surrounded by the quantum phase 
as shown in Section~\ref{sec:offsym}.
In the right panel, the dependence of the expectation value on $\alpha$ is studied for $Q^{SO(2)}(\alpha)$. 
The expectation value is moderately depending on $\alpha$.  As is shown in the histograms in Figure~\ref{fig:aldep},
the system is in the classical phase throughout the region of $\alpha$ in the figure, but the widths of the bunches 
depend on $\alpha$. The widths become the smallest in the middle case. 
This explains the reason why there is a mild peak around $\alpha\sim 0.4$ in the  right panel of Figure~\ref{fig:logairy}.
 
Here it is worth stressing the importance of taking the cosmological constant of CTM to be positive in the discussions 
of this section.
As explained in the last paragraph of Section~\ref{sec:mc}, the Airy function part becomes oscillatory in the positive case, and 
the discussions above can be applied.
On the other hand, when it is negative, the Airy function part becomes a monotonically damping function and the results will 
be different:
In particular, we cannot expect the suppression of the configurations in the quantum phase discussed above, which will highlight 
those in the classical phase.

\begin{figure}
\begin{center}
\includegraphics[width=5cm]{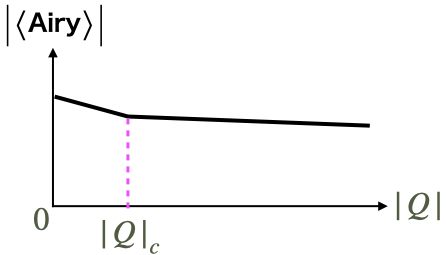}
\hfil
\includegraphics[width=5cm]{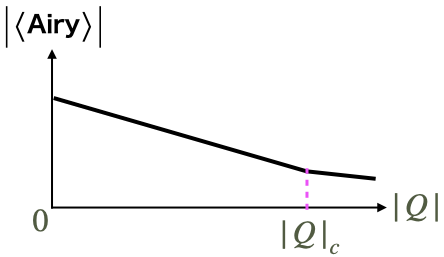}
\hfil
\includegraphics[width=5cm]{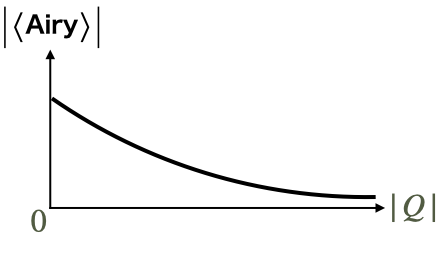}
\caption{Possible three profiles of the behavior of the expectation value. $|Q|_c$ denotes the transition point between
the quantum phase ($|Q|<|Q|_c$) and the classical ($|Q|>|Q|_c$). $|Q|_c$ is smaller in the left panel than that in the middle,
where the former is expected for Lie group invariant $\tilde Q$ and the latter non-invariant, respectively. 
It is even possible that there are no transitions at all as in the right panel.
}
\label{fig:three}
\end{center}
\end{figure}

Lastly let us make some speculations on what happens when we get the dependence of the coupling on $|Q|^2$ back,
namely, $\lambda \rightarrow \lambda |Q|^2$ as in \eq{eq:repsiq}. Now $\lambda |Q|^2$ serves as an effective coupling. 
Then the results from the previous sections tell that, when $|Q|$ is small, the system is in the quantum phase. 
By increasing $|Q|$, the effective coupling becomes larger and the system will eventually undergo the transition to the classical phase 
at some value of $|Q|$, say $|Q|_c$,  which generally depends on $\tilde Q$. 
Note however that there is also the possibility that there are no transitions to the classical phase at all for some $\tilde Q$.  
For Lie group invariant $\tilde Q$, $|Q|_c$ will be relatively smaller,
while $|Q|_c$ will be larger for non-invariant $\tilde Q$. 
Since the expectation value decays more quickly in the quantum phase as $|Q|$ becomes larger, 
we would expect three types of behavior shown in Figure~\ref{fig:three}.
In particular, larger expectation values are expected for Lie group invariant $\tilde Q$ in the large $|Q|$ region.
 
\section{Summary and future prospects}
\label{sec:summary}
In this paper, we have studied the wave function of the canonical tensor model \cite{Sasakura:2011sq}
in the $Q$-representation, when the argument $Q$ takes Lie group invariant or nearby values.
By the Monte Carlo method, 
we have found two phases, which we call the quantum and the classical phases, respectively, where the fluctuations 
of the variables are suppressed in the latter phase than in the former. 
In the classical phase, there emerge configurations which are discretizations of classical geometric spaces 
invariant under the Lie group symmetries:
More precisely,  we have demonstrated the emergence of configurations corresponding to discretized $S^n \, (n=1,2,3)$ 
for  $SO(n+1)$ invariant $Q$.
The transition resembles a matrix model counter part, namely, the transition between one-cut and two-cut solutions in the matrix 
model \cite{Eynard:2016yaa}, 
or the Gross-Witten-Wadia type transition \cite{Gross:1980he,Wadia:1980cp}. 
However, this resemblance is obscured by a difference:
In our setup, the tensor is a given variable for each case 
and the distributions come from the abundance of tensor rank decompositions of the tensor, while,
in the matrix model, the matrix itself has distributions. We have argued that complete splitting of the 
distributions in the classical phase will imply a replica symmetry breaking.  

We have also performed some preliminary studies on the phases and the transition. For larger $N$, the 
transition occurs for smaller $\lambda$, meaning that the classical phase is more favored for larger $N$.
This implies that the emergence of classical spaces is more likely for larger $N$. We have also 
shown some evidences for the importance of Lie group invariance and proper choices of representations 
for the appearance of the classical phase.  

The main difference of the present work from the previous one \cite{Sasakura:2021lub} is the discovery of the classical phase. 
The previous work was not aware of the classical phase, and 
discussed only the quantum phase, and a fluid picture was argued to be useful for 
an approximate analytical treatment for Lie group invariant values of $Q$. 
On the other hand, in the classical phase found in this paper, the inner products $\phi_a^i \phi_a^j$ within the outer bunch 
take almost constant values with small fluctuations. This suggests that a more solid picture than a fluid 
is appropriate to describe the classical phase.
In terms of this aspect, it would be an intriguing coincidence that the form of the real part $Z_{Q,\lambda}$ is similar to that of
 the $p$-spin spherical model for the spin glass \cite{pedestrians}. This suggests that, recalling that the classical phase 
 is accompanied with a replica symmetry breaking, the classical phase would actually be 
 a glassy phase, meaning spacetime in CTM would have similarity to the glass.  
It is an interesting future direction to pursue an approximate analytical method for the classical phase
based on the picture obtained in this paper.

The results of this paper suggest the interesting possibility to apply CTM 
as a model for the evolution of the universe.  As discussed in Section~\ref{sec:oscillatory},
the system is in the quantum phase when $|Q|$ is small, but will eventually undergo the transition 
to the classical phase, when $|Q|$ grows and reaches a critical value depending on $Q/|Q|$, or will never do.
This implies the scenario that the universe starts with the quantum phase with no definite geometry and then enters the classical
phase with the emergence of a space with classical geometry, 
under the assumption that time is correlated with $|Q|$.\footnote{See the footnote \ref{foot:q}.}
It would be a challenging interesting direction of study to compute observable signals, such as primordial fluctuations, 
from the perspective of CTM, and check its significance by comparing with the actual observational data.

Another interesting direction for the application of CTM would be to consider more general values of $Q$ than 
the Lie group invariant ones.
The procedure developed in Section~\ref{sec:trd} to construct $Q$ from geometric data can generally be applied to
a wide range of curved geometries, such as the ones with horizons, singularities, and so on.
Putting these more general values of $Q$ into the argument of the wavefunction and study the properties of the system, such as its phase, 
would provide the general ideas about how spacetime geometries are described in CTM.
In particular, it would be worth studying black hole geometries from the perspective of CTM to shed new light on the long standing 
paradox of information loss.

There are many questions left behind which must be answered in future studies. 
We have only studied the wavefunction for very particular values of the argument, namely, Lie group invariant or nearby ones, 
but these particular choices cannot be well justified without a global view of the wavefunction profile.
We have introduced the parameter $\lambda$ for the convenience of analysis, but the removal limit of taking it infinite 
has not been studied.
We have computed $\langle\prod \hbox{Airy} \rangle_{Q,\lambda}$ but 
not the $Q$ dependence of $Z_{Q,\lambda}$, which is also needed to understand the whole profile of $\Psi(Q,\lambda)$. 
To answer these questions we would need analytical methods, rather than just relying on numerical computations.
The pictures we have obtained so far would provide hints toward them. 

Last but not least, the discovery of the classical phase of this paper suggests a new way of 
tensor rank decomposition \cite{SAPM:SAPM192761164,Carroll1970}.
Tensor rank decomposition is a useful method to extract information from a tensor of real-life data \cite{comon:hal-00923279},
but there is no assurance of getting a good decomposition due to some fundamental problems,
such as the ill-posedness \cite{comon:hal-00923279} of the approximation problem and 
the hardness  \cite{comon:hal-00923279,nphard} of computing the rank of a tensor.
On the other hand, what happens in the classical phase is that  $\phi^i$'s 
are dynamically divided into two bunches, of which the outer bunch serves as a good approximation
of the tensor rank decomposition of $Q$, while the center bunch is a small correction, which could be practically neglected. 
In other words, when we find a classical phase for a tensor $Q$ by taking $\lambda$ large, we can obtain an approximate tensor 
rank decomposition of $Q$.
What seems interesting and useful here is that the rank, namely, the number of $\phi^i$'s in the outer bunch, is automatically
determined by the dynamics of the Monte Carlo simulation, and we do not need to know the rank of $Q$ in advance. 
This would be an advantage considering the hardness of the rank determination, 
though the Monte Carlo simulation is costly.
It would be an interesting question to clarify how useful this method is in practice for various cases of $Q$.

\vspace{.3cm}
\section*{Acknowledgements}
The work of N.S. is supported in part by JSPS KAKENHI Grant No.19K03825. 

\appendix
\def\thesection{Appendix \Alph{section}}

\section{A minimum explanation of CTM}
\label{app:ctm}
This appendix gives a minimum explanation of the canonical tensor model (CTM) to understand the origin of the wave function
discussed in this paper. A longer but concise summary can be found for example in Section 2 of \cite{Obster:2017dhx}
with slightly different normalizations. 

The dynamical variables of CTM are a canonical conjugate pair of real symmetric three-index tensors, $Q_{abc}$ and $P_{abc}$ 
$(a,b,c=1,2,\ldots,N)$.
The quantized variables satisfy
\[
[\hat Q_{abc},\hat P_{def}] =i\sum_{\sigma} \delta_{a \sigma_d}  \delta_{b \sigma_e}  \delta_{a \sigma_f}, 
\]
where the sum is over all the permutations of $d,e,f$.
The expressions of the constraints are given by \cite{Sasakura:2013wza}
\s[
&\hat H_a=\frac{1}{2}\left(
\hat P_{abc}\hat P_{bde}\hat Q_{cde}+ 2 i R \hat P_{abb}-\Lambda_c\, \hat Q_{abb}
\right),\\
&\hat J_{[ab]}=\frac{1}{4}\left(\hat  P_{acd} \hat Q_{bcd}- \hat P_{bcd}\hat Q_{acd}\right),
\s]
where the value of $R$ is determined to be $R=(N+2)(N+3)/4$ from the hermiticity of $\hat H_a$ \cite{Sasakura:2013wza}.   
The square bracket in $\hat J_{[ab]}$ is to represent the anti-symmetry, $\hat J_{[ab]}=-\hat J_{[ba]}$.
$\Lambda_c$ is the cosmological constant \cite{Sasakura:2014gia}, 
and its positivity is essentially important for the appearance of the peak-Lie group relation \cite{Obster:2017dhx,Obster:2017pdq}
mentioned in Section~\ref{sec:introduction}. 
The size of $\Lambda_c$ can be freely changed without loss of generality by performing $(Q,P)\rightarrow (x\,Q, P/x)$ with real $x$.
The expression of the wave function \eq{eq:psip} corresponds to $\Lambda_c=4/9$, which makes the expression convenient 
for later use.

The commutation algebra between the constraints is given by \cite{Sasakura:2013wza,Narain:2014cya}
\s[
&[ \hat H\xi^1,\hat H\xi^2]=i\, \hat J([\hat P\xi^1,\hat P\xi^2]+2 \Lambda_c [\xi^1 ,\xi^2]),\\
&[ \hat J\eta,\, \hat H\xi]=i\,\hat H(\eta \xi),\\
&[ \hat J\eta^1,\,\hat  J\eta^2]=i\,\hat J([\eta^1,\eta^2]),
\label{eq:algctm}
\s]
where we have introduced auxiliary variables $\xi_a,\eta_{[ab]}$ for convenience and 
\s[
&\hat H\xi:=\hat H_a\xi_a,\ \hat J\eta:= \hat J_{[ab]}\eta_{[ab]},\ (\hat P\xi)_{ab}:=\hat P_{abc}\xi_c,\\ 
&[\xi^1,\xi^2]_{ab}:= \xi^1_a \xi^2_b-\xi^1_b \xi^2_a,\ (\eta\xi)_a:= \eta_{[ab]}\xi_b.
\s] 
The algebraic structure \eq{eq:algctm} resembles that of ADM \cite{Arnowitt:1962hi}, and in particular
there exist variable dependent structure coefficients on the righthand side in the first line, which
is also similar to ADM.

\section{Spherical harmonics on $S^d$}
\label{app:spherical}

Let us take a spherical coordinate system $(\varphi,\theta_1,\cdots,\theta_{d-1})$ on $S^d$ with $\varphi\in[-\pi,\pi), \theta_i\in[0,\pi]$. 
If we write $\mathd s_d$ as the line element on $S^d$, $\mathd s_d^2$ can be determined inductively as follows:
\begin{align*}
&\mathd s_1^2=\mathd\varphi^2,\\
&\mathd s_{d+1}^2=\sin^2\theta_d\mathd s_d^2+\mathd\theta_d^2.
\end{align*}
By these relations, the Laplace-Beltrami operator $\Delta_d$ on $S^d$ is also determined inductively as follows:
\begin{align*}
&\Delta_1=\frac{\partial^2}{\partial\varphi^2},\\
&\Delta_{d+1}=\frac{1}{\sin^d\theta_d}\frac{\partial}{\partial\theta_d}\left(\sin^d\theta_d\frac{\partial}{\partial\theta_d}\right)+\frac{\Delta_d}{\sin^2\theta_d}.
\end{align*}
The $d$-dimensional harmonic function on $S^d$ is specified by an array of $d$ integers $L_d=(m,l_1,\cdots,l_{d-1})$ with $|m|\le l_1\le\cdots\le l_{d-1}$, and is written as $Y_{d;L_d}(\varphi,\theta_1,\cdots,\theta_{d-1})$. This satisfies the Helmholtz equation
\begin{align*}
\Delta_d Y_{d;L_d}=-l_{d-1}(l_{d-1}+d-1)Y_{d;L_d}.
\end{align*}
If we assume $Y_{d+1;L_{d+1}}(\varphi,\theta_1,\cdots,\theta_d)=Y_{d;L_d}(\varphi,\theta_1,\cdots,\theta_{d-1})\,\varTheta(\theta_d)$, $\varTheta(\theta_d)$ must be a solution of the equation
\begin{align}
\frac{1}{\sin^d\theta_d}\frac{d}{d\theta_d}\left(\sin^d\theta_d\frac{d \Theta}{d \theta_d}\right)-\frac{l_{d-1}(l_{d-1}+d-1)}{\sin^2\theta_d}\varTheta=-l_d(l_d+d)\varTheta. \label{eq:Theta}
\end{align}
The non-singular solutions of \eqref{eq:Theta} can be written as \cite{gegenbauer}
\begin{align*}
\varTheta(\theta_d)&\propto\frac{1}{\sin^{(d-1)/2}\theta_d}P^{-(l_{d-1}+\frac{d-1}2)}_{l_d+\frac{d-1}2}(\cos\theta_d)\\
&\propto\sin^{l_{d-1}}\theta_d \, C^{l_{d-1}+\frac{d}2}_{l_d-l_{d-1}}(\cos\theta_d),
\end{align*}
where $P^\mu_\nu(x)$ is Ferrers function (or Legendre function) of the first kind 
and $C^\lambda_n(x)$ is Gegenbauer polynomial. They have the relation,
\[
C^\lambda_n(\cos\theta)=\frac{\sqrt{\pi}\, \Gamma(n+2 \lambda)}{2^{\lambda-\frac12} n! \Gamma(\lambda)
}(\sin\theta)^{\frac12-\lambda}P_{n+\lambda-\frac12}^{\frac12-\lambda}(\cos\theta).
\]

This result gives a way to define the spherical harmonics inductively as follows
\begin{align*}
Y_{d+1;L_{d+1}}(\varphi,\theta_1,\cdots,\theta_d)&=
A_{d+1;L_{d+1}}\sin^{l_{d-1}}\theta_d \, C^{l_{d-1}+\frac{d}2}_{l_d-l_{d-1}}(\cos\theta_d)\cdot
Y_{d;L_d}(\varphi,\theta_1,\cdots,\theta_{d-1})
\end{align*}
with a normalization factor $A_{d+1;L_{d+1}}$.
Since the Gegenbauer polynomials satisfy the orthogonality relation \cite{DLMF}
\begin{align*}
\int_{-1}^1 C^\lambda_n(x)C^\lambda_m(x)(1-x^2)^{\lambda-\frac12}\mathd x&=\int_0^\pi C^\lambda_n(\cos\theta)C^\lambda_m(\cos\theta)\sin^{2\lambda}\theta\mathd \theta\\
&=\frac{2^{1-2\lambda}\pi\Gamma(n+2\lambda)}{(n+\lambda)\Gamma(\lambda)^2n!}\delta_{n,m},
\end{align*}
the orthogonality of $Y_{d+1;L_{d+1}}$ is also guaranteed.
Furthermore, the normalization factor $A_{d+1;L_{d+1}}$ is determined by
\begin{align*}
1&=\int_{S^{d+1}}Y_{d+1;L_{d+1}}(\varphi,\theta_1,\cdots,\theta_d)^2\mathd^{d+1}\Omega\\
&=A_{d+1;L_{d+1}}^2\int_0^\pi C^{l_{d-1}+\frac{d}2}_{l_d-l_{d-1}}(\cos\theta_d)^2\sin^{2l_{d-1}+d}\theta_d\mathd\theta_d\\
&=A_{d+1;L_{d+1}}^2\frac{2^{-(2l_{d-1}+d-1)}\pi (l_d+l_{d-1}+d-1)!}{(l_d+\frac{d}2)\Gamma(l_{d-1}+\frac{d}2)^2(l_d-l_{d-1})!}.
\end{align*}
This shows the specific value of $A_{d+1;L_{d+1}}$:
\begin{align*}
A_{d+1;L_{d+1}}=\left(\frac{(l_d+\frac{d}2)\Gamma(l_{d-1}+\frac{d}2)^2(l_d-l_{d-1})!}{2^{-(2l_{d-1}+d-1)}\pi(l_d+l_{d-1}+d-1)!}\right)^{1/2}.
\end{align*}
Collecting the results above, we obtain \eq{eq:genY}.


\vspace{.3cm} 
\begin{thebibliography}{}

\bibitem{Georgi:1999wka}
H.~Georgi,
``Lie algebras in particle physics,''
Front. Phys. \textbf{54}, 1-320 (1999).

\bibitem{Reuter:2019byg} 
  M.~Reuter and F.~Saueressig,
  ``Quantum Gravity and the Functional Renormalization Group : The Road towards Asymptotic Safety,'' Cambridge University Press, 2019.

\bibitem{Eichhorn:2018yfc}
A.~Eichhorn,
``An asymptotically safe guide to quantum gravity and matter,''
Front. Astron. Space Sci. \textbf{5}, 47 (2019)
doi:10.3389/fspas.2018.00047
[arXiv:1810.07615 [hep-th]].

 
\bibitem{Loll:2019rdj}
R.~Loll,
``Quantum Gravity from Causal Dynamical Triangulations: A Review,''
Class. Quant. Grav. \textbf{37}, no.1, 013002 (2020)
doi:10.1088/1361-6382/ab57c7
[arXiv:1905.08669 [hep-th]].
  
\bibitem{Rovelli:2014ssa} 
  C.~Rovelli and F.~Vidotto,
  ``Covariant Loop Quantum Gravity : An Elementary Introduction to Quantum Gravity and Spinfoam Theory,''
  Cambridge University Press, 2014.
  
\bibitem{Surya:2019ndm}
S.~Surya,
``The causal set approach to quantum gravity,''
Living Rev. Rel. \textbf{22}, no.1, 5 (2019)
doi:10.1007/s41114-019-0023-1
[arXiv:1903.11544 [gr-qc]].
  
\bibitem{Konopka:2006hu} 
  T.~Konopka, F.~Markopoulou and L.~Smolin,
  ``Quantum Graphity,''
  hep-th/0611197.
  
\bibitem{Wolfram:2020jjc}
S.~Wolfram,
``A Class of Models with the Potential to Represent Fundamental Physics,''
Complex Syst. \textbf{29}, 107-536 (2020)
doi:10.25088/ComplexSystems.29.2.107
[arXiv:2004.08210 [cs.DM]].

\bibitem{Trugenberger:2016viw}
C.~A.~Trugenberger,
``Combinatorial Quantum Gravity: Geometry from Random Bits,''
JHEP \textbf{09}, 045 (2017)
doi:10.1007/JHEP09(2017)045
[arXiv:1610.05934 [hep-th]].

\bibitem{Akara-pipattana:2021zzy}
P.~Akara-Pipattana, T.~Chotibut and O.~Evnin,
``The birth of geometry in exponential random graphs,''
[arXiv:2102.11477 [cond-mat.dis-nn]].


\bibitem{Ambjorn:1990ge}
J.~Ambjorn, B.~Durhuus, and T.~Jonsson, ``{Three-dimensional simplicial quantum
  gravity and generalized matrix models},''
\href{http://dx.doi.org/10.1142/S0217732391001184}{{\em Mod. Phys. Lett.}
  {\bfseries A06} (1991) 1133--1146}.

\bibitem{Sasakura:1990fs}
N.~Sasakura, ``{Tensor model for gravity and orientability of manifold},''
\href{http://dx.doi.org/10.1142/S0217732391003055}{{\em Mod. Phys. Lett.}
  {\bfseries A06} (1991) 2613--2624}.

\bibitem{Godfrey:1990dt}
N.~Godfrey and M.~Gross, ``{Simplicial quantum gravity in more than
  two-dimensions},''
\href{http://dx.doi.org/10.1103/PhysRevD.43.1749}{{\em Phys. Rev.} {\bfseries
  D43} (1991) R1749--1753}.

\bibitem{Gurau:2009tw} 
  R.~Gurau,
  ``Colored Group Field Theory,''
  Commun.\ Math.\ Phys.\  {\bf 304}, 69 (2011)
  doi:10.1007/s00220-011-1226-9
  [arXiv:0907.2582 [hep-th]].

\bibitem{Sasakura:2011sq}
N.~Sasakura,
``Canonical tensor models with local time,''
Int. J. Mod. Phys. A \textbf{27}, 1250020 (2012)
doi:10.1142/S0217751X12500200
[arXiv:1111.2790 [hep-th]].

\bibitem{Bonzom:2011zz}
V.~Bonzom, R.~Gurau, A.~Riello and V.~Rivasseau,
``Critical behavior of colored tensor models in the large N limit,''
Nucl. Phys. B \textbf{853}, 174-195 (2011)
doi:10.1016/j.nuclphysb.2011.07.022
[arXiv:1105.3122 [hep-th]].
  
\bibitem{Gurau:2011xp}
R.~Gurau and J.~P.~Ryan,
``Colored Tensor Models - a review,''
SIGMA \textbf{8}, 020 (2012)
doi:10.3842/SIGMA.2012.020
[arXiv:1109.4812 [hep-th]].


\bibitem{Sasakura:2012fb}
N.~Sasakura,
``Uniqueness of canonical tensor model with local time,''
Int. J. Mod. Phys. A \textbf{27}, 1250096 (2012)
doi:10.1142/S0217751X12500960
[arXiv:1203.0421 [hep-th]].

\bibitem{Arnowitt:1962hi}
R.~L.~Arnowitt, S.~Deser and C.~W.~Misner,
``The Dynamics of general relativity,''
Gen. Rel. Grav. \textbf{40}, 1997-2027 (2008)
doi:10.1007/s10714-008-0661-1
[arXiv:gr-qc/0405109 [gr-qc]].

\bibitem{Ambjorn:2004qm}
J.~Ambjorn, J.~Jurkiewicz and R.~Loll,
``Emergence of a 4-D world from causal quantum gravity,''
Phys. Rev. Lett. \textbf{93}, 131301 (2004)
doi:10.1103/PhysRevLett.93.131301
[arXiv:hep-th/0404156 [hep-th]].

\bibitem{Obster:2017dhx}
D.~Obster and N.~Sasakura,
``Emergent symmetries in the canonical tensor model,''
PTEP \textbf{2018}, no.4, 043A01 (2018)
doi:10.1093/ptep/pty038
[arXiv:1710.07449 [hep-th]].

\bibitem{Obster:2017pdq}
D.~Obster and N.~Sasakura,
``Symmetric configurations highlighted by collective quantum coherence,''
Eur. Phys. J. C \textbf{77}, no.11, 783 (2017)
doi:10.1140/epjc/s10052-017-5355-y
[arXiv:1704.02113 [hep-th]].


\bibitem{Sasakura:2021lub}
N.~Sasakura,
``Phase profile of the wave function of canonical tensor model and emergence of large spacetimes,''
[arXiv:2104.11845 [hep-th]].

\bibitem{Eynard:2016yaa}
B.~Eynard,
``Counting Surfaces,''
Prog.~Math.~Phys. 70 (2016)
doi:10.1007/978-3-7643-8797-6

\bibitem{Gross:1980he}
D.~J.~Gross and E.~Witten,
``Possible Third Order Phase Transition in the Large N Lattice Gauge Theory,''
Phys. Rev. D \textbf{21}, 446-453 (1980)
doi:10.1103/PhysRevD.21.446

\bibitem{Wadia:1980cp}
S.~R.~Wadia,
``$N$ = Infinity Phase Transition in a Class of Exactly Soluble Model Lattice Gauge Theories,''
Phys. Lett. B \textbf{93}, 403-410 (1980)
doi:10.1016/0370-2693(80)90353-6

\bibitem{Sasakura:2020rqz}
N.~Sasakura,
``Symmetry enhancement in a two-logarithm matrix model and the canonical tensor model,''
PTEP \textbf{2021}, no.4, 043A01 (2021)
doi:10.1093/ptep/ptab034
[arXiv:2008.07726 [hep-th]].


\bibitem{Sasakura:2013wza}
N.~Sasakura,
``Quantum canonical tensor model and an exact wave function,''
Int. J. Mod. Phys. A \textbf{28}, 1350111 (2013)
doi:10.1142/S0217751X1350111X
[arXiv:1305.6389 [hep-th]].

\bibitem{DeWitt:1967yk}
B.~S.~DeWitt,
``Quantum Theory of Gravity. 1. The Canonical Theory,''
Phys. Rev. \textbf{160}, 1113-1148 (1967)
doi:10.1103/PhysRev.160.1113

\bibitem{Narain:2014cya}
G.~Narain, N.~Sasakura and Y.~Sato,
``Physical states in the canonical tensor model from the perspective of random tensor networks,''
JHEP \textbf{01}, 010 (2015)
doi:10.1007/JHEP01(2015)010
[arXiv:1410.2683 [hep-th]].


\bibitem{Sasakura:2014gia}
N.~Sasakura and Y.~Sato,
``Interpreting canonical tensor model in minisuperspace,''
Phys. Lett. B \textbf{732}, 32-35 (2014)
doi:10.1016/j.physletb.2014.03.006
[arXiv:1401.2062 [hep-th]].

\bibitem{Witten:2010cx}
E.~Witten,
``Analytic Continuation Of Chern-Simons Theory,''
AMS/IP Stud. Adv. Math. \textbf{50}, 347-446 (2011)
[arXiv:1001.2933 [hep-th]].

\bibitem{Obster:2020vfo}
D.~Obster and N.~Sasakura,
``Phases of a matrix model with non-pairwise index contractions,''
PTEP \textbf{2020}, no.7, 073B06 (2020)
doi:10.1093/ptep/ptaa085
[arXiv:2004.03152 [hep-th]].

\bibitem{Obster:2021xtb}
D.~Obster and N.~Sasakura,
``Counting Tensor Rank Decompositions,''
Universe \textbf{7}, no.8, 302 (2021)
doi:10.3390/universe7080302
[arXiv:2107.10237 [gr-qc]].

\bibitem{Berger:2019odf}
C.~E.~Berger, L.~Rammelm\"uller, A.~C.~Loheac, F.~Ehmann, J.~Braun and J.~E.~Drut,
``Complex Langevin and other approaches to the sign problem in quantum many-body physics,''
Phys. Rept. \textbf{892}, 1-54 (2021)
doi:10.1016/j.physrep.2020.09.002
[arXiv:1907.10183 [cond-mat.quant-gas]].


\bibitem{Neal(2011)} 
R.~Neal, Handbook of Markov Chain Monte Carlo, 113 (2011). 
doi:10.1201/b10905
[arXiv:1206.1901 [stat.CO]].

\bibitem{paralleltemp}
 D.~J.~Earl and M.~W.~Deem,  
 ``Parallel tempering: Theory, applications, and new perspectives". Phys. Chem. Chem. Phys. 7: 3910 (2005) doi:10.1039/b509983h.


\bibitem{boost} \url{https://www.boost.org}

\bibitem{Lionni:2019rty}
L.~Lionni and N.~Sasakura,
``A random matrix model with non-pairwise contracted indices,''
PTEP \textbf{2019}, no.7, 073A01 (2019)
doi:10.1093/ptep/ptz057
[arXiv:1903.05944 [hep-th]].

\bibitem{Sasakura:2019hql}
N.~Sasakura and S.~Takeuchi,
``Numerical and analytical analyses of a matrix model with non-pairwise contracted indices,''
Eur. Phys. J. C \textbf{80}, no.2, 118 (2020)
doi:10.1140/epjc/s10052-019-7591-9
[arXiv:1907.06137 [hep-th]].

\bibitem{Kawano:2018pip}
T.~Kawano, D.~Obster and N.~Sasakura,
``Canonical tensor model through data analysis: Dimensions, topologies, and geometries,''
Phys. Rev. D \textbf{97}, no.12, 124061 (2018)
doi:10.1103/PhysRevD.97.124061
[arXiv:1805.04800 [hep-th]].


\bibitem{gegenbauer}
H.~S.~Cohl, T.~H.~Dang and T.~M.~Dunster,
 ``Fundamental solutions and Gegenbauer expansions of Helmholtz operators in Riemannian spaces of constant curvature'', 
SIGMA \textbf{14} (2018), 136 
doi.org/10.3842/SIGMA.2018.136
[ arXiv:1803.07149v2].  



\bibitem{HFT}
S.~Axler, ``HFT.m'', 
\url{https://www.axler.net/HFT_Math.html}


\bibitem{Hartnoll:2019pwe}
S.~A.~Hartnoll, E.~A.~Mazenc and Z.~D.~Shi,
``Topological order in matrix Ising models,''
SciPost Phys. \textbf{7}, no.6, 081 (2019)
doi:10.21468/SciPostPhys.7.6.081
[arXiv:1908.07058 [hep-th]].


\bibitem{SAPM:SAPM192761164}
F.~L. Hitchcock, ``The expression of a tensor or a polyadic as a sum of
  products,'' \href{http://dx.doi.org/10.1002/sapm192761164}{{\em Journal of
  Mathematics and Physics} {\bfseries 6} no.~1-4, (1927) 164--189}.
  \url{http://dx.doi.org/10.1002/sapm192761164}.

\bibitem{Carroll1970}
J.~D. Carroll and J.-J. Chang, ``Analysis of individual differences in
  multidimensional scaling via an n-way generalization of ``eckart-young''
  decomposition,'' \href{http://dx.doi.org/10.1007/BF02310791}{{\em
  Psychometrika} {\bfseries 35} no.~3, (Sep, 1970) 283--319}.
  \url{https://doi.org/10.1007/BF02310791}.

\bibitem{Landsberg2012}
{Landsberg, J. M.}, {\em {Tensors: Geometry and Applications}}.
\newblock {American Mathematical Society, Providence}, {2012}.

\bibitem{comon:hal-00923279}
P.~Comon, ``{Tensors: a Brief Introduction},''
  \href{http://dx.doi.org/10.1109/MSP.2014.2298533}{{\em {IEEE Signal
  Processing Magazine}} {\bfseries 31} no.~3, (May, 2014) 44--53}.
  \url{https://hal.archives-ouvertes.fr/hal-00923279}.
  
\bibitem{pedestrians}
T.~Castellani and A.~Cavagna, ``Spin-glass theory for pedestrians'', 
J.~Stat.~Mech.: Theo.~Exp. {\bf 2005}, 05012
[arXiv: cond-mat/0505032].  

\bibitem{qi1}
G.~Ni, L.~Qi, F.~Wang, Y.~Wang, ``The degree of the E-characteristic polynomial of an even order tensor", 
J.~Math.~Anal.~Appl. 329 (2007) 1218-1229.

\bibitem{qi2}
L.~Qi, ``Eigenvalues and invariants of tensors", J. Math. Anal. Appl. 325 (2007) 1363-1377.

\bibitem{tensoreigen}
D.~Cartwright and B.~Sturmfels, ``The number of eigenvalues of a tensor",
Linear~Alg.~Appl. 438:2 (2013) 942-952, 
[arXiv:1004.4953v2], 
\url{https://doi.org/10.1016/j.laa.2011.05.040}.


 \bibitem{nphard}  
C.~J.~Hillar and L.~Lim, ``Most tensor problems are NP-Hard", Journal of the ACM. 60 (6) (2013): 1-39. arXiv:0911.1393. doi:10.1145/2512329

\bibitem{DLMF}
NIST Digital Library of Mathematical Functions, 
http://dlmf.nist.gov/, 
Release 1.1.2 of 2021-06-15, 
F.~W.~J. Olver, A.~B. {Olde Daalhuis}, D.~W. Lozier, B.~I. Schneider, R.~F. Boisvert, C.~W. Clark, B.~R. Miller, B.~V. Saunders, H.~S. Cohl, and M.~A. McClain, eds.


\end{thebibliography}
\end{document}